\documentclass[twocolumn,preprintnumbers,showpacs,aps]{revtex4}
\usepackage{epic,eepic}
\usepackage{hyperref}

\begin{document}

\newcommand{\lp}{\left}
\newcommand{\rp}{\right}
\newcommand{\be}{\begin{eqnarray}}
\newcommand{\ee}{\end{eqnarray}}
\newcommand{\beq}{\begin{equation}}
\newcommand{\eeq}{\end{equation}}
\newcommand{\ba}{\begin{array}}
\newcommand{\ea}{\end{array}}

\newcommand{\braket}[2]{{\langle{#1}|{#2}\rangle}}
\newcommand{\ket}[1]{{|{#1}\rangle}}
\newcommand{\bra}[1]{{\langle{#1}|}}
\newcommand{\av}[1]{\langle#1\rangle}
\newcommand{\bbraket}[2]{{\left\langle{#1}|{#2}\right\rangle}}
\newcommand{\bket}[1]{{\left|{#1}\right\rangle}}
\newcommand{\bbra}[1]{{\left\langle{#1}\right|}}
\newcommand{\bav}[1]{\left\langle#1\right\rangle}


\preprint{CALT-68-2393}

\title{Anyons from non-solvable finite groups\\
are sufficient for universal quantum computation}

\author{Carlos Mochon}
\email{carlosm@theory.caltech.edu}
\affiliation{Institute for Quantum Information, 
California Institute of Technology,
Pasadena, CA 91125,USA}

\date{March 3, 2003}

\begin{abstract}
We present a constructive proof that anyonic magnetic charges with fluxes in 
a non-solvable finite group can perform universal quantum computations.
The gates are built out of the elementary operations of braiding, fusion,
and vacuum pair creation, supplemented by a reservoir of ancillas
of known flux. Procedures for building the ancilla reservoir and
for correcting leakage are also described. 
Finally, a universal qudit gate-set, which is ideally
suited for anyons, is presented. The gate-set
consists of classical computation supplemented by 
measurements of the $X$ operator.
\end{abstract}



\maketitle


\section{Introduction}

The discovery of the potential speedups offered by quantum computers
launched an effort to find physical systems out of which these 
computers could be built. Researchers soon found that these systems
are in short supply, as it is extremely difficult to isolate a quantum 
system from the environment, while maintaining enough control to perform 
operations on the encoded data.
The advent of quantum error correction and fault-tolerant processing
has drastically increased the tolerable error rates; nonetheless,
physical systems with low enough error rates are still hard to come by.

One way to protect a Hilbert space from the environment is to 
encode the quantum data in non-local observables. These observables,
which are constructed from topological invariants, cannot be measured or
changed by any local operator. Therefore, because the environment generally
acts locally, the physics of the system provides a form of fault tolerance.

In particular, consider the spectrum of electrically and 
magnetically charged particles that are obtained by breaking a
gauge group to a finite subgroup.
The finite group gauge theory is a particularly good system for
quantum computing because it involves no gauge fields, and hence
no long-range interactions except for those obtained by braiding.
Furthermore, the Hamiltonian of the system respects the unbroken
symmetry; therefore Schur's lemma forbids the types of coupling
to uncharged objects that can produce decoherence. Of course, 
the data could still decay by the exchange of a charged particle
between two anyons, but this is a quantum tunneling event which is
exponentially suppressed by the distance between the particles.

When the gauge theory is restricted to two spatial dimensions,
the particles acquire topological long-range interactions,
which can be be used to perform computations. The interactions occur
when the particles are exchanged or braided, and depend only on
the topological class of the path involved. 
Because of these interactions, the charged particles  
have quantum statistics that are more exotic than the standard fermions 
and bosons, and are known as anyons. The non-standard statistics,
though, only arise when clockwise rotations 
can be distinguished from counterclockwise rotations, which is why we 
impose the requirement of two spatial dimensions.
While this two-dimensional model of the world seems somehow unphysical,
there exist condensed-matter systems with quasiparticles that behave like
anyons.

The original proposal for an anyon based quantum computer was made by 
Kitaev \cite{Kitaev:1997wr,Freedman:2001}. 
The first concrete description was done by Ogburn and Preskill 
in Refs.~\cite{Ogburn:1998,Preskill:1997uk} for anyons in the
group $A_5$, the even permutations of five elements.
In our paper, we will generalize the work of Ogburn and Preskill
to any non-solvable finite group, which includes $A_5$ as 
the smallest case.

The paper is organized as follows:  We begin by introducing some notation 
and reviewing the properties of the anyon model that will be used throughout 
this paper. 
Section \ref{sec:gates} presents the universal gate-set that will 
be employed to prove anyons can perform quantum computations. 
Sections \ref{sec:simp} through \ref{sec:hard} contain the meat
of the paper, and discuss a concrete anyonic implementation of all the
necessary gates. For pedagogical reasons, we first cover the easier subcase
of simple perfect groups in Section~\ref{sec:simp}, and then discuss 
the required generalizations for any non-solvable group in 
Section~\ref{sec:hard}. In Section~\ref{sec:leak}, we discuss how to make 
these computations fault-tolerant by performing leakage correction. Finally, 
we discuss the conclusions and unsolved questions. There are also two 
appendices which include the mathematical proofs, and a technique for creating
anyon ancillas.

\section{Review}

In this section we will review some of the braiding and fusion properties
of our anyons. Our review will be rather abridged, but more details can
be found in the excellent review of discrete gauge theories 
\cite{deWildPropitius:1995hk} (and the original work
\cite{Bais:1992pe}). The paper by Ogburn and Preskill 
\cite{Ogburn:1998,Preskill:1997uk}
also contains a good review with emphasis on the applications to quantum
computing.

This section also establishes our notation for qudits, and reviews
the phase estimation circuit, a highly useful trick that will be used 
often.

\subsection{\label{sec:flux}Magnetic charges}

The main players throughout this paper will be the magnetic charges, also 
known as fluxes. For a field theory with an unbroken finite
group $G$, there is one magnetic charge for each element $g\in G$.
Quantum mechanically, we can have superpositions of these states,
giving a one-particle Hilbert space spanned by $\ket{g}$ for all $g\in G$
(though strictly speaking, superpositions of charges in different conjugacy 
classes are meaningless, as will be explained in the next subsection).

\begin{figure}
\setlength{\unitlength}{0.00083333in}
{\renewcommand{\dashlinestretch}{30}
\begin{picture}(2188,1541)(0,-10)
\dashline{60.000}(725,1262)(12,1262)
\dashline{60.000}(1438,1262)(2151,1262)
\dashline{60.000}(725,447)(12,447)
\dashline{60.000}(1438,447)(2151,447)
\put(1794.500,448.000){\arc{508.001}{3.7831}{5.6417}}
\blacken\path(1925.179,641.826)(1998.000,600.000)(1953.433,671.176)(1925.179,641.826)
\put(1794.500,446.000){\arc{508.001}{0.6415}{2.5001}}
\blacken\path(1663.821,252.174)(1591.000,294.000)(1635.567,222.824)(1663.821,252.174)
\put(1794.500,1262.000){\arc{508.001}{3.7831}{5.6417}}
\blacken\path(1635.567,1485.176)(1591.000,1414.000)(1663.821,1455.826)(1635.567,1485.176)
\put(1794.500,1261.000){\arc{508.001}{0.6415}{2.5001}}
\blacken\path(1953.433,1037.824)(1998.000,1109.000)(1925.179,1067.174)(1953.433,1037.824)
\put(1998,447){\blacken\ellipse{144}{144}}
\put(1998,447){\ellipse{144}{144}}
\put(165,447){\blacken\ellipse{144}{144}}
\put(165,447){\ellipse{144}{144}}
\put(572,447){\blacken\ellipse{144}{144}}
\put(572,447){\ellipse{144}{144}}
\put(1591,447){\blacken\ellipse{144}{144}}
\put(1591,447){\ellipse{144}{144}}
\put(1998,1262){\blacken\ellipse{144}{144}}
\put(1998,1262){\ellipse{144}{144}}
\put(165,1262){\blacken\ellipse{144}{144}}
\put(165,1262){\ellipse{144}{144}}
\put(572,1262){\blacken\ellipse{144}{144}}
\put(572,1262){\ellipse{144}{144}}
\put(1591,1262){\blacken\ellipse{144}{144}}
\put(1591,1262){\ellipse{144}{144}}
\thicklines
\path(929,447)(1285,447)
\blacken\path(1183.160,406.265)(1285.000,447.000)(1183.160,487.735)(1213.712,447.000)(1183.160,406.265)
\path(929,1262)(1285,1262)
\blacken\path(1183.160,1221.265)(1285.000,1262.000)(1183.160,1302.735)(1213.712,1262.000)(1183.160,1221.265)
\put(165,192){\makebox(0,0)[b]{$h$}}
\put(572,192){\makebox(0,0)[b]{$g$}}
\put(165,1007){\makebox(0,0)[b]{$h$}}
\put(572,1007){\makebox(0,0)[b]{$g$}}
\put(1998,830){\makebox(0,0)[b]{$h$}}
\put(1591,830){\makebox(0,0)[b]{$h g h^{-1}$}}
\put(1591,40){\makebox(0,0)[b]{$g\ $}}
\put(1998,40){\makebox(0,0)[b]{$\, g^{-1}h g$}}
\end{picture}
}
\caption{\label{switch}Exchanging two anyons.}
\end{figure}
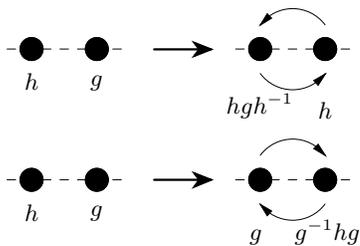

Specifying the exchange properties of the charges involves
making a choice of gauge. The easiest choice, which will be used in
this paper, is to keep all anyons ordered on a horizontal line.
The exchange of particles, which can be clockwise or counterclockwise, 
is only allowed between adjacent pairs. In either case, the particle
that passes below remains unchanged, while the particle that passes above
gets conjugated. When the exchange is in the counterclockwise direction,
the upper anyon gets conjugated by the flux of the lower one, whereas
in the clockwise direction it gets conjugated by the inverse of the lower
flux. This is depicted in Figure~\ref{switch}.

One way to visualize these exchanges is to associate with each anyon
a ray that is vertical in the plane, starting at the particle and proceeding 
upwards.
Anyons are allowed to move freely through the plane, but every time an 
anyon crosses the ray of another particle, it gets conjugated
by the flux of the owner of the ray (or by the inverse flux if crossing from
left to right). Note that when a particle passes a group of anyons,
it gets conjugated by the total flux of the anyons, which is given as the 
product from left to right of the individual fluxes.

Clearly, moving single anyons around can produce strange correlations
throughout the system. However, moving a pair of anyons with
a total flux that is trivial will not change the state of the 
system if the pair always passes below. This is why we will always be dealing 
with states of the form
\be
\sum_g a_g \ket{g}\otimes \ket{g^{-1}},
\ee
\noindent
which correspond to a pair of anyons with trivial total flux. When dealing
only with pairs of trivial total flux, we can swap any two pairs,
or bring any two pairs together without affecting the state of the
rest of the system.

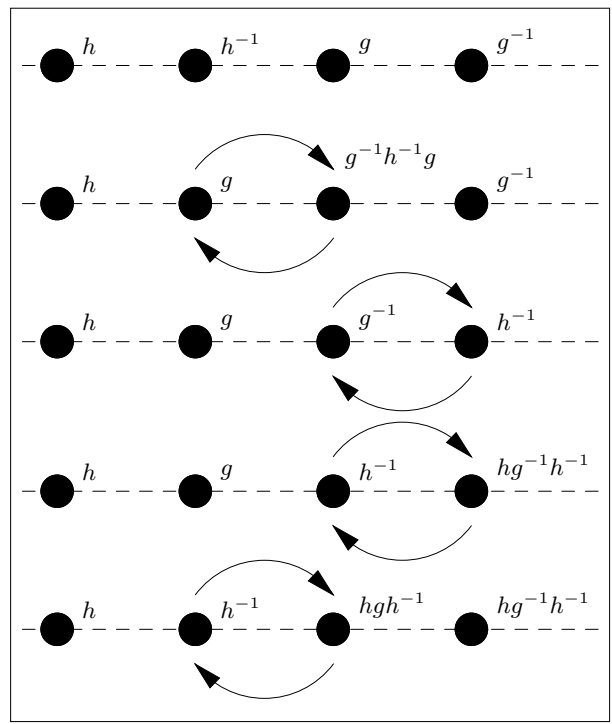
\begin{figure}
\setlength{\unitlength}{0.0012in}
{\renewcommand{\dashlinestretch}{30}
\begin{picture}(2624,3139)(0,-10)
\dashline{60.000}(62,2262)(2562,2262)
\dashline{60.000}(62,1012)(2562,1012)
\dashline{60.000}(62,412)(2562,412)
\dashline{60.000}(62,2862)(2562,2862)
\dashline{60.000}(62,1662)(2562,1662)
\put(1112.000,2337.000){\arc{750.000}{0.6435}{2.4981}}
\blacken\path(919.374,2050.593)(812.000,2112.000)(877.845,2007.288)(919.374,2050.593)
\put(1712.000,1737.000){\arc{750.000}{0.6435}{2.4981}}
\blacken\path(1519.374,1450.593)(1412.000,1512.000)(1477.845,1407.288)(1519.374,1450.593)
\put(1712.000,937.000){\arc{750.000}{3.7851}{5.6397}}
\blacken\path(1904.626,1223.407)(2012.000,1162.000)(1946.155,1266.712)(1904.626,1223.407)
\put(1712.000,1087.000){\arc{750.000}{0.6435}{2.4981}}
\blacken\path(1519.374,800.593)(1412.000,862.000)(1477.845,757.288)(1519.374,800.593)
\put(1112.000,337.000){\arc{750.000}{3.7851}{5.6397}}
\blacken\path(1304.626,623.407)(1412.000,562.000)(1346.155,666.712)(1304.626,623.407)
\put(1112.000,487.000){\arc{750.000}{0.6435}{2.4981}}
\blacken\path(919.374,200.593)(812.000,262.000)(877.845,157.288)(919.374,200.593)
\put(1112.000,2187.000){\arc{750.000}{3.7851}{5.6397}}
\blacken\path(1304.626,2473.407)(1412.000,2412.000)(1346.155,2516.712)(1304.626,2473.407)
\put(1712.000,1587.000){\arc{750.000}{3.7851}{5.6397}}
\blacken\path(1904.626,1873.407)(2012.000,1812.000)(1946.155,1916.712)(1904.626,1873.407)
\put(1412,412){\blacken\ellipse{142}{142}}
\put(1412,412){\ellipse{142}{142}}
\put(2012,412){\blacken\ellipse{142}{142}}
\put(2012,412){\ellipse{142}{142}}
\put(212,412){\blacken\ellipse{142}{142}}
\put(212,412){\ellipse{142}{142}}
\put(812,412){\blacken\ellipse{142}{142}}
\put(812,412){\ellipse{142}{142}}
\put(1412,1012){\blacken\ellipse{142}{142}}
\put(1412,1012){\ellipse{142}{142}}
\put(2012,1012){\blacken\ellipse{142}{142}}
\put(2012,1012){\ellipse{142}{142}}
\put(212,1012){\blacken\ellipse{142}{142}}
\put(212,1012){\ellipse{142}{142}}
\put(812,1012){\blacken\ellipse{142}{142}}
\put(812,1012){\ellipse{142}{142}}
\put(1412,1662){\blacken\ellipse{142}{142}}
\put(1412,1662){\ellipse{142}{142}}
\put(2012,1662){\blacken\ellipse{142}{142}}
\put(2012,1662){\ellipse{142}{142}}
\put(212,1662){\blacken\ellipse{142}{142}}
\put(212,1662){\ellipse{142}{142}}
\put(812,1662){\blacken\ellipse{142}{142}}
\put(812,1662){\ellipse{142}{142}}
\put(1412,2262){\blacken\ellipse{142}{142}}
\put(1412,2262){\ellipse{142}{142}}
\put(2012,2262){\blacken\ellipse{142}{142}}
\put(2012,2262){\ellipse{142}{142}}
\put(212,2262){\blacken\ellipse{142}{142}}
\put(212,2262){\ellipse{142}{142}}
\put(812,2262){\blacken\ellipse{142}{142}}
\put(812,2262){\ellipse{142}{142}}
\put(1412,2862){\blacken\ellipse{142}{142}}
\put(1412,2862){\ellipse{142}{142}}
\put(2012,2862){\blacken\ellipse{142}{142}}
\put(2012,2862){\ellipse{142}{142}}
\put(212,2862){\blacken\ellipse{142}{142}}
\put(212,2862){\ellipse{142}{142}}
\put(812,2862){\blacken\ellipse{142}{142}}
\put(812,2862){\ellipse{142}{142}}
\path(12,3112)(2612,3112)(2612,12)
	(12,12)(12,3112)
\put(322,2912){\makebox(0,0)[lb]{$h$}}
\put(922,2912){\makebox(0,0)[lb]{$h^{-1}$}}
\put(1522,2912){\makebox(0,0)[lb]{$g$}}
\put(2122,2912){\makebox(0,0)[lb]{$g^{-1}$}}
\put(322,2312){\makebox(0,0)[lb]{$h$}}
\put(922,2312){\makebox(0,0)[lb]{$g$}}
\put(1462,2412){\makebox(0,0)[lb]{$g^{-1}h^{-1}g$}}
\put(2122,2312){\makebox(0,0)[lb]{$g^{-1}$}}
\put(322,1712){\makebox(0,0)[lb]{$h$}}
\put(922,1712){\makebox(0,0)[lb]{$g$}}
\put(1522,1712){\makebox(0,0)[lb]{$g^{-1}$}}
\put(2122,1712){\makebox(0,0)[lb]{$h^{-1}$}}
\put(322,1062){\makebox(0,0)[lb]{$h$}}
\put(922,1062){\makebox(0,0)[lb]{$g$}}
\put(1522,1062){\makebox(0,0)[lb]{$h^{-1}$}}
\put(2122,1062){\makebox(0,0)[lb]{$h g^{-1} h^{-1}$}}
\put(322,462){\makebox(0,0)[lb]{$h$}}
\put(922,462){\makebox(0,0)[lb]{$h^{-1}$}}
\put(1522,462){\makebox(0,0)[lb]{$h g h^{-1}$}}
\put(2122,462){\makebox(0,0)[lb]{$h g^{-1} h^{-1}$}}
\end{picture}
}
\caption{\label{braid}Conjugating a pair of anyons.}
\end{figure}

We do want to allow controlled interaction between pairs, though, and
this is accomplished by a pass-through operation. The idea is to have one
pair circle one anyon from the other pair. This will conjugate the
fluxes of the pair that circles, but leave the other pair invariant.
This operation is depicted using elementary exchanges in 
Figure~\ref{braid}.

The net result of the pictured operation is 
\be
&&\ket{h}\otimes\ket{h^{-1}}\otimes\ket{g}\otimes\ket{g^{-1}}
\nonumber\\ &&\longrightarrow
\ket{h}\otimes\ket{h^{-1}}\otimes\ket{h g h^{-1}}\otimes\ket{h g^{-1}h^{-1}}
,
\ee
\noindent
which is a conjugation of the second pair by $h$. Conjugation by $h^{-1}$
could be achieved by using counterclockwise exchanges in the
picture.

For notational convenience, in this paper we will generally only 
mention the flux of the left element of a pair. The above transformation
will be written as
\be
\ket{h}\otimes\ket{g}
\longrightarrow
\ket{h}\otimes\ket{h g h^{-1}},
\ee
\noindent
leaving the compensating fluxes implicit. While we will exclusively
deal in this paper with flux pairs with trivial flux, we will only 
explicitly refer to the second anyon when necessary to describe the 
operations.

\subsection{\label{sec:elec}Electric charges and vacuum pairs}

We now wish to focus on the operations of creating pairs from the vacuum
and fusing pairs back into the vacuum. However, we must first 
briefly discuss the complete spectrum of particles, and that involves 
electric charges.

An electric charge is a particle with no flux that transforms as some
non-trivial irreducible representation of the group $G$. A useful analogy
is to think of the representation of $G$ as the total spin of the particle.
The internal state of the particle is then equivalent to the direction in
which the spin is pointing.

The electric charge states can be labeled as $\ket{R, V}$, where $R$
is a representation of $G$ and $V$ is a vector that transforms in the
representation $R$. The electric charges do not interact with each other, but
when one of them circles a magnetic flux $g$, its state changes as
\be
\ket{R, V} \longrightarrow \ket{R, U_R(g) V},
\ee
\noindent
where $U_R(g)$ is the matrix corresponding to $g$ in the representation $R$.
This is known as the Aharonov-Bohm effect.

While we can transform the state of an electric charge within the subspace
of a representation, there are no operations (other than fusion, which
destroys the particle) that can change the
representation of a particle. Furthermore, the phase
between states of different representations cannot be measured.
We can therefore effectively describe the electric charges 
as having decohered into the different representations. 
In particle physics we would say that the different
representations correspond to different superselection sectors.

The same thing happens to the magnetic charges. Different conjugacy
classes live in different superselection sectors, so we can imagine
that there is an automatic decoherence into different conjugacy
classes. Superpositions of fluxes in different conjugacy
classes are therefore meaningless.

The spectrum also contains particles with both electric
and magnetic charge, which are called dyons. The only special feature 
is that the electric charge is a representation only of the subgroup of 
$G$ that commutes with the flux. The aforementioned magnetic charges
are just dyons with a trivial representation. The dyons also have
superselection sectors that correspond to different conjugacy classes and
representations.

The purpose of discussing the full spectrum, and the idea of superselection
sectors, is to find out what kind of states we get when we
create a pair of particles from the vacuum. The first thing to note
is that each of the particles will instantly decohere into a specific 
conjugacy class and representation. Furthermore, because a pair created
from the vacuum must have trivial total charge and flux, the conjugacy
classes must be inverses, and the representations must be conjugate
representations.

Consider the case that the pair decoheres into plain magnetic charges,
with the first one contained in the conjugacy class $C$.
Because the combined state still has vacuum quantum numbers,
the state must transform trivially when another flux is dragged around it.
That is, it must be invariant under conjugation. There is only one
such state:
\be
\ket{\text{Vac}\lp(C\rp)} = 
\frac{1}{\sqrt{\lp|C\rp|}}
\sum_{g\in C} \ket{g} \otimes \ket{g^{-1}}.
\ee
\noindent
The vacuum states for the other superselection sectors are also unique
and have similar expressions. When a pair of anyons is created from the 
vacuum, it will start off in one of these states.

Another useful operation is to fuse two anyons together. Note that
we are not talking about two anyon pairs, but rather two anyons, sometimes
from the same pair, and sometimes from different pairs.
The operation of fusion will turn the two particles into one, which must
carry the total flux and charge of the two. It is also possible that the
two anyons will have vacuum quantum numbers, and will fuse back into the
vacuum. In this case, no particle will be left behind and their
total mass will be transformed into some other medium, such as radiation.
If $\ket{\Psi}$ is the combined state of the two anyons, and the first 
anyon is in the conjugacy class $C$, then the
probability that the two will fuse into the vacuum is given by
the standard rules of quantum mechanics:
\be
P_{vacuum} = \lp| \braket{\text{Vac}\lp(C\rp)}{\Psi} \rp|^2.
\ee

After fusing two particles of different pairs, the fused particle may
carry some flux. However, since the total flux of the original four 
particles was trivial, the total flux of all the remaining particles
(including the product of the fusion) 
will be trivial as well. Therefore, it is possible to safely move
the group of particles away from the bulk of the computation without
disturbing our quantum state.

\subsection{Qudits}

Throughout this paper it will be useful to perform computations
with qudits rather than the usual qubits. We define our computational
basis as the states $\ket{i}$ for $0\leq i < d$, where we will assume
that $d$ is prime. The unitary $Z$ and $X$ gates can be defined as follows
\be
Z \ket{i} &=& \omega^{i} \ket{i},\\
X \ket{i} &=& \ket{i+1},
\ee
\noindent
where $\omega$ is a fixed non-trivial $d^{th}$ root of unity, and sums
are understood to be modulo $d$. The operators satisfy the commutation 
relation
\be
Z X = X Z \omega.
\ee

As usual, the eigenstates of $Z$ correspond to the computational basis.
We can also introduce the eigenstates of $X$:
\be
\ket{\tilde i} = \frac{1}{\sqrt{d}} \sum_{j=0}^{d-1} \omega^{-i j} \ket{j},
\ee
\noindent
which have the following transformations under the action of our unitary
gates:
\be
Z \ket{\tilde i} &=& \ket{\widetilde{i-1}},\\
X \ket{\tilde i} &=& \omega^{i} \ket{\tilde i}.
\ee

\subsection{Phase measurement}

A very useful trick, used many times throughout this paper, is 
Kitaev's phase estimation technique \cite{Kitaev:1995qy}. 
In fact, we will only employ
a special case of the technique which we describe below.

Assume that we are working in a system with qudits, and we have an operator
$U$ with eigenvalues that are $d^{th}$ roots of unity. We shall prove that 
being able to apply a controlled-$U$, and measure in the $X$ basis,
is equivalent to being able to measure the operator $U$.

Consider applying the circuit below to an eigenstate $\ket{\Psi_j}$
of $U$ with eigenvalue $\omega^j$:

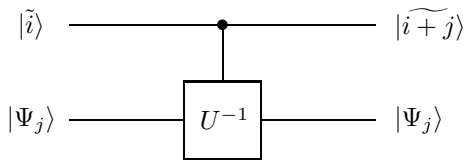
\begin{center}
\setlength{\unitlength}{1in}
\begin{picture}(3,1)(0,0)
{\put(.7,.8){\line( 1, 0){1.6}}}
{\put(.7,.3){\line( 1, 0){.6}}}
{\put(2.3,.3){\line( -1, 0){.6}}}
{\put(1.5,.8){\line( 0,-1){.3}}}
{\thinlines \put(1.5,.8){\circle*{.05}}}
{\put(1.3,.1){\framebox(.4,.4){}}}

\put(1.51,.31){\makebox(0,0){$U^{-1}$}}
\put(.5,.8){\makebox(0,0){$\ket{\tilde i}$}}
\put(.5,.3){\makebox(0,0){$\ket{\Psi_j}$}}
\put(2.4,.8){\makebox(0,0)[l]{$|\widetilde{i+j}\rangle$}}
\put(2.4,.3){\makebox(0,0)[l]{$\ket{\Psi_j}$}}
\end{picture}
\end{center}

\noindent
where the controlled-$U^{-1}$ can be applied as $d-1$ 
controlled-$U$'s.
The circuit works as described because the controlled-$U^{-1}$
leaves the bottom state invariant, but applies a $Z^{-j}$ to the
upper state. On a general state $\ket{\phi} = \sum_j c_j \ket{\Psi_j}$ 
expanded in terms of eigenvectors of $U$, the circuit produces the 
transformation
\be
\ket{\tilde 0}\otimes \ket{\phi} 
\longrightarrow \sum_j c_j \ket{\tilde j}\otimes\ket{\Psi_j}.
\ee
\noindent
Clearly, a subsequent measurement of the first qudit in the $X$ basis is 
equivalent to a non-destructive measurement of the original state in the 
$U$ basis.
We will use this technique in the next section to measure the operators 
$X^a Z^b$. 

In a later section we will employ the equivalent circuit

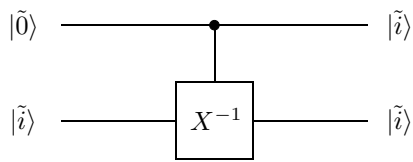
\begin{center}
\setlength{\unitlength}{1in}
\begin{picture}(3,1)(0,0)
{\put(.7,.8){\line( 1, 0){1.6}}}
{\put(.7,.3){\line( 1, 0){.6}}}
{\put(2.3,.3){\line( -1, 0){.6}}}
{\put(1.5,.8){\line( 0,-1){.3}}}
{\thinlines \put(1.5,.8){\circle*{.05}}}
{\put(1.3,.1){\framebox(.4,.4){}}}

\put(1.51,.31){\makebox(0,0){$X^{-1}$}}
\put(.5,.8){\makebox(0,0){$\ket{\tilde 0}$}}
\put(.5,.3){\makebox(0,0){$\ket{\tilde i}$}}
\put(2.4,.8){\makebox(0,0)[l]{$\ket{\tilde i}$}}
\put(2.4,.3){\makebox(0,0)[l]{$\ket{\tilde i}$}}
\end{picture}
\end{center}

\noindent
run in both the forwards and backwards direction, to change 
between the $\ket{\tilde i}$ states and the readily available
$\ket{\tilde 0}$ state which can be naturally produced from, 
and fused into the vacuum.

\section{\label{sec:gates}A Universal Gate-set for Anyons}

A lot of the work in proving universality can be simplified by choosing a 
proper gate-set. For this paper, we will employ a generalization of the 
gate-set used by Ogburn and Preskill \cite{Ogburn:1998,Preskill:1997uk}. 
The gate-set,
which involves measurements as well as unitary gates,
can be applied to qudits when $d$ is prime, which is the only case
considered in this paper.

The universal gate-set is: 
\begin{enumerate}
\item Measure non-destructively $Z$
\item Measure non-destructively $X$
\item Apply Toffoli operators 
(to any set of three qudits)
\end{enumerate}
\noindent
where the qudit Toffoli is defined as
\be
T \ket{l,m,n} = \ket{l,m,l m+n}.
\ee
\noindent
and all computations are done modulo $d$. 

Though tangential to the main purpose of this paper, the above 
gate-set is another answer to the question posed in Ref.~\cite{Shi:2002}.
That is, given a Toffoli, what extra gates are required to complete
a universal set? Of course, the answer provided by the above gate-set
involves measurements in an integral way, and is therefore different from
the one proposed in Ref.~\cite{Shi:2002}. However, the above gate-set
also addresses the question:
Given classical computation (i.e. Toffoli and measurements of $Z$), what
gates are needed to complete the universal set?

We now turn our attention to the proof of universality for the gate-set
presented above. We note that Gottesman has already
proven in Ref.~\cite{Gottesman:1998se} that for $d$ prime, applying and 
measuring products of
$Z$'s and $X$'s, plus a Toffoli, is universal for quantum computation. 
All we need to do in order to prove universality, is to show that we can 
apply and measure operators of the form $X^a Z^b$ using the above gates.

Measurements of $X$ followed by measurements of $Z$ can produce $\ket{i}$
ancillas for any $i$. Similarly, we can obtain $\ket{\tilde i}$ ancillas from
measurements of $Z$ followed by measurements of $X$. A controlled-sum
can be made out of a Toffoli by fixing an input to a $\ket{1}$ ancilla. 
Because a controlled-sum is really a controlled-$X$, fixing the other input to 
$\ket{1}$ produces the $X$ gate. On the other hand, a controlled-sum
from a state to a $\ket{\tilde 1}$ ancilla, produces a $Z$ on the state:

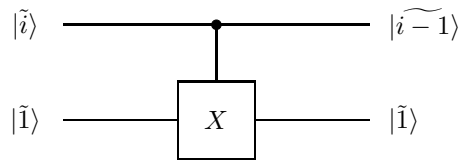
\begin{center}
\setlength{\unitlength}{1in}
\begin{picture}(3,1)(0,0)
{\put(.7,.8){\line( 1, 0){1.6}}}
{\put(.7,.3){\line( 1, 0){.6}}}
{\put(2.3,.3){\line( -1, 0){.6}}}
{\put(1.5,.8){\line( 0,-1){.3}}}
{\thinlines \put(1.5,.8){\circle*{.05}}}
{\put(1.3,.1){\framebox(.4,.4){}}}

\put(1.5,.3){\makebox(0,0){$X$}}
\put(.5,.8){\makebox(0,0){$\ket{\tilde i}$}}
\put(.5,.3){\makebox(0,0){$\ket{\tilde 1}$}}
\put(2.4,.8){\makebox(0,0)[l]{$|\widetilde{i-1}\rangle$}}
\put(2.4,.3){\makebox(0,0)[l]{$\ket{\tilde 1}$}}
\end{picture}
\end{center}

The general case of applying $X^a Z^b$ can be done by a series of $X$ 
and $Z$ gates. All that remains is to construct a method for measuring 
operators of the form $X^a Z^b$. First, we note that 
\be
\lp(X^a Z^b\rp)^d &=& \omega^{a b d(d-1)/2} X^{ad} Z^{b d} \nonumber \\
&=& \lp\{ \matrix{1& &d\  \text{odd} \cr -1^{a b}& &d=2.} \rp.
\ee

\subsection{$d$ odd case}

The case $d=2$ is rather complicated and will be handled separately. The 
general case $d$ odd (remember we required $d$ prime) is easy because
the eigenvalues of $X^a Z^b$ are the $d^{th}$ roots of unity just like those
of $X$ and $Z$. As discussed in the review of phase estimation, 
being able to apply
a controlled-$X^a Z^b$, combined with measurements in the $X$ basis
(which includes preparation of $X$ eigenstates) is sufficient to measure
in the $X^a Z^b$ basis. 

All that remains is to construct the controlled-$X^a Z^b$. That is, we
need to be able to apply the gate
\be
\ket{n} \otimes \ket{\psi} &\longrightarrow& \ket{n} 
\otimes \lp(X^a Z^b\rp)^n 
\ket{\psi} \nonumber \\
&=& \ket{n} \otimes X^{an} Z^{b n} \omega^{a b n(n-1)/2} \ket{\psi},
\ee
\noindent
composed of a phase
$\ket{n,m} \rightarrow \omega^{b n m + a b n(n-1)/2} \ket{n,m}$
followed by controlled-sums. The controlled-sum is just a Toffoli with an 
input fixed to one.
As for the phase, because we have a Toffoli, we have universal classical 
computation. We can thus compute $b n m + a b n(n-1)/2$ in an ancilla, apply
a $Z$ to this ancilla, and then erase the computation. 

\subsection{$d=2$ case}

The $d=2$ case is somewhat trickier because our gate-set is invariant under 
complex conjugation, and thus there is no way of distinguishing the two
eigenstates of $ZX=iY$. We will solve this problem by creating an
ancilla that is an eigenstate of $ZX$, defining it to be the $+i$ 
eigenstate, and then using it to measure and build more eigenstates.

Assume we were given a state 
\be
\ket{\Psi} = \frac{1}{\sqrt{2}} \lp(\ket{0} + \omega \ket{1}\rp),
\ee
\noindent
where $\omega^2 = -1$. Clearly, the state is equal to one of the
two $ZX$ eigenstates: $\ket{\pm_Y} = \lp(\ket{0} \pm i\ket{1}\rp)/\sqrt{2}$. 

Using a controlled-$Z X$, which is built by the method described in the
$d$ odd case, we can produce copies of the state $\ket{\Psi}$. The idea,
similar to the one used for phase estimation, is to apply the 
controlled-$Z X$ from a state $\ket{\tilde 0}$ to the state $\ket{\Psi}$.
The target state is an eigenvector of $Z X$ with eigenvalue $\omega$,
and therefore the relative phase is copied over to the first state:
\be
\ket{\tilde 0} \otimes \ket{\Psi} &\longrightarrow& \nonumber
\frac{1}{\sqrt{2}} \lp( \ket{0} + \omega \ket{1} \rp) \otimes \ket{\Psi} \\
&=& \ket{\Psi} \otimes\ket{\Psi}.
\ee
\noindent
Notice that copying works independently of whether $\ket{\Psi}$ 
is the $+i$ or $-i$ eigenstate of $ZX$. Naturally, by subsequently applying 
a $Z$, we can also produce the orthogonal state 
$\ket{\Phi} = \lp(\ket{0} - \omega \ket{1}\rp)/\sqrt{2}$.

With our ancilla, we can also measure in this basis. This is done by applying
a controlled-$Z X$ to the ancilla from the state we want to measure:
\be
\ket{\Psi} \otimes \ket{\Psi} &\longrightarrow&
\ket{\tilde 1} \otimes \ket{\Psi},\\
\ket{\Phi} \otimes \ket{\Psi} &\longrightarrow&
\ket{\tilde 0} \otimes \ket{\Psi},
\ee
\noindent
and then measuring in the $X$ basis.

As long as we are consistent in always using the same ancilla $\ket{\Psi}$, 
we will have broken the conjugation symmetry, and found a way to label, 
create and measure eigenstates of $ZX$. Of course, we should keep many copies 
of the ancilla, which can be prepared from the original state.
The operations above also allow us to error correct
our set of ancillas by copying each, comparing the copies, and using
majority voting to discard the damaged ancillas. Thus, even if there are
some errors in preparation, or some of the ancillas decay over time, 
computation will still be feasible.

All that remains to be explained is how to create the first copy of 
$\ket{\Psi}$.
Because a state with a density-matrix proportional to the identity
can be written as
\be
\rho = \frac{1}{2} I =
\frac{1}{2} \ket{+_Y} \bra{+_Y}   + \frac{1}{2} \ket{-_Y} \bra{-_Y},
\ee
\noindent
it is equivalent to having prepared an eigenstate of $ZX = iY$ 
chosen at random. The state $\rho=I/2$ can be produced by discarding
one qubit of a bell state, and a bell state can be produced with a 
controlled-sum from a $\ket{\tilde 0}$ ancilla to a $\ket{0}$ ancilla.
Therefore, we have shown that we can produce the initial eigenstate of 
$Z X$, and we have completed the proof that the gate-set presented at
the beginning of this section is universal for quantum computation.

\section{\label{sec:simp}Universal Computation for Simple Perfect groups}

In this section we will prove that a set of anyons based on certain groups
can perform universal quantum computations. Instead of dealing first with
the general case of non-solvable groups, we will deal with the smaller 
set of groups that are both simple and perfect.

We remind the reader that non-solvable groups are those that contain a
perfect subgroup; and a perfect group is any non-trivial group, whose 
commutator subgroup equals the full group: $[G,G]=G$. 
The property of simplicity means that
the group has exactly two subgroups that are invariant under conjugation:
the trivial group and the whole group. Because the commutator subgroup
is invariant under conjugation, it should be clear that any simple
non-abelian group is perfect. However, we shall refer to these groups
as simple and perfect to remind the reader that we are dealing with
a subcase of the general non-solvable case. 

The set of simple perfect groups, which includes the groups $A_n$ for $n>4$, 
is powerful 
for computing because in some sense we can get from one non-trivial element 
to any other using operations that fix the identity. The general case of 
non-solvable groups will be deferred to Section~\ref{sec:hard}, 
where we will show that a simple perfect group can be extracted from 
a non-solvable group.

\subsection{\label{sec:reqs}Requirements for the physical system}

Here we list the operations, ancillas, and measurements that we assume
are available on any realistic anyonic system, and which we will use
to build our quantum gate-set:
\begin{enumerate}
\item We can braid or exchange any two particles.
\item We can fuse a pair of anyons and detect whether there is a particle
left behind or whether they had vacuum quantum numbers.
\item We can produce a pair of anyons in a state that is chosen at
random from the two particle subspace that has vacuum quantum numbers.
\item We have ancilla pairs of the form $\ket{g}\otimes\ket{g^{-1}}$
for any $g\in G$, where the individual anyons have trivial electric charge.
\end{enumerate}

We remind the reader again that even though all our anyons
are used in pairs of trivial total flux, we will generally only mention 
one of the anyons of the pair. These conventions also apply to ancillas,
which means that we will refer to the $\ket{g}\otimes\ket{g^{-1}}$ state
as an ancilla of flux $g$.

While the first three requirements are natural operations for a laboratory
system, it is not clear where the ancillas would come from. Depending on 
the physical realization there may be many ways of obtaining the ancilla
reservoir. We discuss one such scheme in Appendix~\ref{sec:ancillas}.

\subsection{\label{sec:basis}Computational basis}

Let $G$ be a simple and perfect finite group. Let $a$ and $b$ be
two non-commuting elements of $G$. Let $d$ be the smallest integer such that
$a^d b a^{-d} = b$. We can assume that $d$ is prime, otherwise we could replace
$a$ by $a^{d/p}$ where $p$ is some prime that divides $d$.

It turns out that every simple non-abelian group has even order.
This was first conjectured by Burnside \cite{Burnside:1911} in 1911, and
proven by Feit and Thompson \cite{Feit:1963} in 1963
(in fact, the complete classification of simple finite groups was completed
in the early 1980's, see for instance Ref.~\cite{Gorenstein:1994}).
Using Sylow's theorems, the fact that every simple group has even order means
that they all include a non-trivial element $a$ such that $a^2=1$. Therefore,
we could always work with a basis of qubits. However, we will present
the general qudit case both for its elegance, and because in some instances
a basis of qudits is more convenient.

We will work with a basis of qudits of trivial net flux
\be
\ket{n} = \ket{a^n b a^{-n}} \otimes \ket{a^n b^{-1} a^{-n}}
\ee
\noindent
for $0\leq n<d$, where we have explicitly described both
anyons of the pair. 

It should be clear that we can initialize the computer by filling up the
computational space with $\ket{0}$ ancillas. We turn now to the task
of constructing the gates presented in Section~\ref{sec:gates}.

\subsection{Conjugation by a function}

We begin by describing the technique of conjugation by a function, which is 
especially powerful for simple perfect groups. In Section~\ref{sec:flux}
we showed that we could perform the transformation
\be
\ket{h}\otimes\ket{g}
\longrightarrow
\ket{h}\otimes\ket{h g h^{-1}},
\ee
\noindent
where we conjugate the second anyon by the flux of the first, while
the first anyon remains invariant. We can also conjugate an anyon
by a product $h_1 h_2\cdots h_n$ 
\be
\ket{g}
\longrightarrow
\ket{h_1 h_2 \cdots h_n g h_n^{-1} \cdots h_2^{-1} h_1^{-1}},
\ee
\noindent
where the $\{h_i\}$ are fluxes of other anyons which remain unchanged 
throughout this process. The procedure is done by first conjugating by $h_n$, 
then by $h_{n-1}$, and proceeding leftward until we finally conjugate
by $h_1$.

The above procedure is not terribly useful if all the $\{h_i\}$ are fluxes
of fixed ancillas, because we could have equivalently conjugated
by a single ancilla of flux $h=h_1 h_2\cdots h_n$. However, some of the
fluxes in the product could correspond to anyons that represent qudits of 
unknown state. In this case we can think of the above operation as conjugation
by a function of the fluxes of certain qudits. 

Let us consider what kind of functions can be applied in this way. Clearly 
we are speaking about functions that can be written as products of elements
of $G$. The elements can include known constants if we use our ancillas to
conjugate. We can also include  the flux of a qudit, which will be of the
form $a^i b a^{-i}$ if the qudit is in the computational basis (though
this may not
be the case when we are trying to correct leakage). Finally, we can 
include in the product the inverse of the flux of a qudit, as discussed in 
Section~\ref{sec:flux}.

In conclusion, given $n$ qudits with fluxes $g_1$
through $g_n$, and a function $f(g_1,\dots,g_{n-1})$ of the first $n-1$
qudits, we can conjugate the last qudit by $f$
\be
\ket{g_{n}} \longrightarrow 
\ket{f(g_1,\dots,g_{n-1}) g_{n} f(g_1,\dots,g_{n-1})^{-1}},
\ee
\noindent
provided that the function $f$ can be written in product form.
By product form, we mean that $f$ is a product of the inputs
$\{g_i\}$, their inverses $\{g_i^{-1}\}$, and fixed elements
of $G$, each of which may appear more than once, or not at all. 
For example, a valid function would be 
$f(g_1,g_2) = a g_2 b g_1^{-1} c g_1^{-1} d$ with $a,b,c,d \in G$.
Furthermore, this transformation does not change the flux of the first
$n-1$ qudits, though it may entangle them with the last qudit.

\subsection{\label{sec:toff}Toffoli gate}

To build the Toffoli gate we must be able to conjugate the third qudit
by the function $f(g_1,g_2)$, which depends on the fluxes of the first
two qudits as
\be
f(a^i b a^{-i},a^j b a^{-j})=a^{i j},
\ee
\noindent
and is arbitrary for values of $g_1$ and $g_2$ that are not 
in the computational basis. If the third qudit is in the state $a^k b a^{-k}$,
conjugation by $f$ produces the transformation
\be
\ket{a^k b a^{-k}} \longrightarrow \ket{a^{i j + k} b a^{-i j -k}},
\ee
\noindent
which is the desired Toffoli gate.

Given the discussion in the previous subsection, we are left with the task of 
expressing the function $f$ in product form. However, it turns out that 
for simple and perfect groups every function has such an expression:

\textbf{Theorem:} 
If $G$ is a simple and perfect finite group, then any 
function $f(g_1,\dots,g_n):G^n\rightarrow G$ can be expressed as a product of 
the inputs $\{g_i\}$, their inverses $\{g_i^{-1}\}$ and fixed elements of $G$, 
any of which may appear multiple times in the product.

Not only does the above theorem prove that Toffoli gates are possible
for any simple and perfect group, but it directly proves that any classical 
function can be computed.

The proof of the theorem, which is mostly constructive, is somewhat long and 
will be deferred to Appendix \ref{sec:proofs}. However, to make this seem 
plausible to the casual reader, we would like to illustrate the basic steps 
needed to build a Toffoli gate for qubits.

The main idea behind the construction is that the function $f$ is basically
a logical \textsc{and} of the inputs. A commutator makes a good logical
\textsc{and} because it equals the identity if either of its inputs are the 
identity.
Furthermore, the commutator function can be expanded as a product of 
its inputs.  Therefore, we would like the first input to take values 
$1$ or $c$ and the second input to take values $1$ or $d$, with the requirement
that $d$ not commute with $c$, so that we can put them into a commutator.

Let $g_1$ denote the flux of the first qubit, and $g_2$ the flux of the
second qubit. Each takes values $g_i\in\lp\{b,a b a^{-1}\rp\}$. Define
the new variables $g_i' = g_i b^{-1} \in \lp\{1,c\rp\}$, where 
$c \equiv [a,b] \equiv a b a^{-1} b^{-1}$. 
It is sufficient to show that we can express the Toffoli function
as a product of $g_1'$, $g_2'$, their inverses and fixed ancillas. 

Choose an element $d$ that does not commute with $c$ and define
$e \equiv [c,d]$. Imagine we could find two functions of one element,
that can be expressed in product form, such that
\be
h_1(c) = d&&\ \ \ h_1(1)=1,\\
h_2(e) = a&&\ \ \ h_2(1)=1.
\ee

Using these functions, the Toffoli function can be written as
\be
f(g_1,g_2) &=& h_2\bigg(\ \bigg[g_1',\  h_1(g_2')\bigg]\ \bigg),
\ee
\noindent
which when expanded out is a product of the correct form.

The existence of the functions $h_i$, which is discussed in more detail
in the full proof of the theorem, is a consequence of $G$ being simple.
For any element $c\in G$, the group generated by its conjugacy class $C(c)$
is a normal subgroup. Because $G$ is simple, this subgroup must equal the full 
group. Therefore, every element $d\in G$ has an expression of the form
$d = x_1 c x_1^{-1} x_2 c x_2^{-1} \cdots  x_n c x_n^{-1}$ for some
$n$ and some elements $\{x_i\}\in G$. We can use the expression to construct
$h_1$:
\be
h_1(g) = x_1 g x_1^{-1} x_2 g x_2^{-1} \cdots  x_n g x_n^{-1},
\ee
\noindent
and a similar construction builds $h_2$.

For a concrete example we can work with $G=A_5$. We begin by choosing an
element $a$, which must satisfy $a^2=I$, if we wish to work with qubits
($d=2$). Because of the symmetries of the group, all choices are equivalent 
to $a = (12)(34)$. The next step would involve choosing an element
$b$ that does not commute with $a$, and an element $d$ that does not
commute with $c\equiv[a,b]$. While any choice can produce a Toffoli, the
required $h_1$ function will be simplified if we can make
$c$ and $d$ fall in the same conjugacy class. The same can be said
for $h_2$ if $e\equiv[c,d]$ and $a$ are in the same conjugacy class.

At this point, a little trial and error yield $b = (345)$ and
$d = (234)$. The computational basis is now defined as
\be
\ket{0} &=& \ket{b} = \ket{(345)},\nonumber \\
\ket{1} &=& \ket{a b a^{-1}} = \ket{(435)},
\ee
\noindent
and the remaining group elements are fixed as
\be
c &=& \lp(a b a^{-1}\rp) b^{-1} = (435) (435) = (345), \nonumber \\
e &=& \lp(c d c^{-1}\rp) d^{-1} = (245) (324) = (25)(34).
\ee
\noindent
The $h_i$ functions, which are the only non-constructive part of the proof,
can be built as simple conjugations because of the choices we made earlier:
\be
h_1(g) = h_2(g) = (521) g (125),
\ee
\noindent
where both happen to be the same function by coincidence.
Putting all the steps together we have a function
\be
&f\lp(g_1,g_2\rp)& \! = \bigg \{(521)\bigg 
         [ g_1 (435), (521) g_2 (435) (125) \bigg ](125)\bigg\}
\nonumber\\
&=& \! \! \! \! \! \bigg\{ (521) g_1 (435)      (521) g_2 (435)      (125) 
\nonumber\\*
&&                         (345) g_1^{-1} (521) (345) g_2^{-1} (125) (125) 
\bigg\} \nonumber\\
&=& \! \! \! \! \! \bigg\{ (521) g_1 (14352) g_2 (124) 
g_1^{-1} (15342) g_2^{-1} (521) \bigg\} \nonumber
\ee
\noindent
which can be applied with nine elementary conjugations.

\subsection{Measuring $Z$}

The basic idea behind measuring in the computational basis is that if
we fuse a flux with another flux of the inverse group element, there is
a finite chance that they will have vacuum quantum numbers and disappear.
On the other hand, if the product of the two fluxes is not unity then there 
must be a particle left behind to carry the remaining flux (i.e., the total
flux is always conserved).

At this point it might be useful to remind the reader why a fusion of
$g$ with $g^{-1}$ will not always turn into the vacuum. The short story is
that the combined state is not invariant when another flux encircles them,
implying that they have an electric charge component. The state that has vacuum
quantum numbers is invariant under the effect of all
fluxes, and hence is the sum of all the states in the conjugacy class of $g$, 
with the same phase. We can figure out the probability of fusion into the 
vacuum by calculating the overlap of the vacuum state with 
the state of two anyons to be fused. The result is 
\be
P = \lp| \bra{\text{Vac}\lp(C\rp)} 
         \lp( \ket{g} \otimes \ket{g^{-1}} \rp)\rp|^2 = 
\frac{1}{\lp|C(g)\rp|}
\ee
\noindent
where $C(g)$ is the conjugacy class of $g$, and the vacuum state was
defined in Section~\ref{sec:elec}.

Because one fusion will only probabilistically tell us the desired result,
we should repeat the measurement many times to obtain a sufficient degree
of accuracy. Besides, if we are working with qudits with $d>2$ we need to test 
fusion with at least two different fluxes. We therefore need to have many 
copies of the state to be measured. 

Because of the no cloning theorem, copying cannot be done exactly, but the 
transformation
\be
\sum_i C_i \ket{i} \longrightarrow \sum_i C_i 
\ket{i} \otimes  \ket{i} \otimes \ket{i} \otimes \cdots \otimes \ket{i}
\ee
\noindent
means that we can measure each of the separate copies in the $Z$ basis
and expect to get the same answer. The above transformation can be done
with a controlled-sum (Toffoli with one input fixed to $\ket{1}$) 
from the original state
to a $\ket{0}$ ancilla. Repeating this controlled-sum
with many target ancillas will produce the above entangled state.

To summarize, the procedure for measuring in the $Z$ basis is first to
create an entangled state using a controlled-sum. Then try to fuse each
of the qudits with one of the inverses of the fluxes that are $Z$ eigenstates.
Eventually, one will disappear into the vacuum, and the inverse of the flux
of that ancilla is the result. Even in the presence of errors, this 
measurement will have a good fidelity because the probability of failure is 
exponentially small in the number of fusions.

A final note is that, because we are always dealing with pairs of fluxes,
what fusion really means is that we fuse the first anyon of our qubit
with the first anyon of the ancilla.

\subsection{Constructing the zero eigenvector of $X$}

For the next gates, we are going to need a supply of states that are
eigenvectors of $X$ with zero eigenvalue:
\be
\ket{\tilde 0} = \frac{1}{\sqrt d} \sum_{i=0}^{d-1} \ket{i}.
\ee

We will produce them out of pairs of anyons with vacuum quantum numbers.
As usual we will just discuss one member of the pair, and assume that
the equivalent operations are being performed on the other anyon.

One of the possible states that (when paired) have vacuum quantum numbers
is the sum of fluxes in the conjugacy class of $b$. This is approximately
what we want. Sadly, in general, a state created from the vacuum will be a 
mix of this desired state plus other states, including states
that involve dyonic particles (particles with both electric and magnetic 
charge). We will have to filter through all this noise to get our $X$ 
eigenstate.

The procedure that we will describe below is effectively an incomplete
swap, that has been extended to the full Hilbert space in a logical
way. In the computation basis, the operations act as

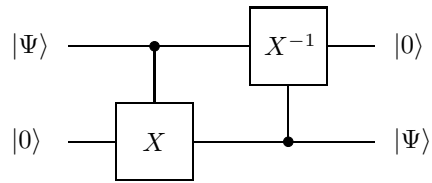
\begin{center}
\setlength{\unitlength}{1in}
\begin{picture}(3,1.1)(0,0)
{\put(.7,.8){\line( 1, 0){.95}}}
{\put(2.3,.8){\line( -1, 0){.25}}}
{\put(.7,.3){\line( 1, 0){.25}}}
{\put(2.3,.3){\line( -1, 0){.95}}}

{\put(1.15,.8){\line( 0,-1){.3}}}
{\thinlines \put(1.15,.8){\circle*{.05}}}
{\put(0.95,.1){\framebox(.4,.4){}}}

{\put(1.85,.3){\line( 0,1){.3}}}
{\thinlines \put(1.85,.3){\circle*{.05}}}
{\put(1.65,.6){\framebox(.4,.4){}}}

\put(1.15,.3){\makebox(0,0){$X$}}
\put(1.86,.81){\makebox(0,0){$X^{-1}$}}
\put(.4,.8){\makebox(0,0)[l]{$\ket{\Psi}$}}
\put(.4,.3){\makebox(0,0)[l]{$\ket{0}$}}
\put(2.4,.8){\makebox(0,0)[l]{$\ket{0}$}}
\put(2.4,.3){\makebox(0,0)[l]{$\ket{\Psi}$}}
\end{picture}
\end{center}

\noindent
which performs a swap provided that the second qudit started in the 
$\ket{0}$ state. Outside of the computational basis, though, the operations
are chosen so that we can detect whether we obtained the desired 
$\ket{\tilde 0}$ state or not.

We start with two qudit states, one created from the vacuum and one
which is a $\ket{0}$ ancilla:
\be
\ket{\text{Vac}} \otimes \ket{0} = \lp(C \ket{\tilde 0} 
+ D \ket{\Psi_\perp}\rp) \otimes \ket{0},
\ee
\noindent
where $\ket{\Psi_\perp}$ is a state orthogonal to the computational
subspace. If the vacuum pair decohered into a superselection sector
other than the one that contains the computational basis, the constant
$C$ will be zero. This will not be a problem as we will be able
to detect this case, and then start again from this step.

Using the theorem from Section~\ref{sec:toff}, 
we can conjugate the $\ket{0}$ ancilla by a
function of the flux of the vacuum pair that has the following form:
\be
f(a^i b a^{-i}) &=& a^i,\nonumber\\*
f(\text{anything else}) &=& I,
\ee
\noindent
which is essentially a controlled-sum that has been properly defined outside
the computational basis.

The state of the combined system after conjugation will be 
\beq
\frac{C}{\sqrt{d}} \sum_{i=0}^{d-1} \ket{a^i b a^{-i}} 
\otimes \ket{a^i b a^{-i}} + 
\sum_{i=0}^{d-1} D_{i} \ket{\Psi_{i\perp}} \otimes \ket{a^i b a^{-i}}, 
\eeq
\noindent
where $\{D_i\}$ are some constants, and $\{\ket{\Psi_{i\perp}}\}$
are states perpendicular to the computational basis. 
Note that the states $\ket{\Psi_{i\perp}}$ 
for $i\neq 0$ are 
the ones that have flux $a^i b a^{-i}$ but have non-trivial charge. 
The state $\ket{\Psi_{0\perp}}$ includes all the other fluxes and charges.
Depending on the superselection sector in which the vacuum state was 
created, many of the constants $C$ and $\{D_i\}$ will be zero.

Now we conjugate by $f^{-1}$ from the ancilla to the vacuum state yielding
\be
C \ket{b} \otimes \ket{\tilde 0} + 
\sum_{i=0}^{d-1} D_{i} \ket{\Psi_{i\perp}'} \otimes \ket{a^i b a^{-i}},
\ee
\noindent
where $\{\ket{\Psi_{0\perp}'}\}=\{\ket{\Psi_{0\perp}}\}$ and the states
$\{\ket{\Psi_{i\perp}'},i>0\}$ have flux $b$ but non-trivial charge.

Now we try to fuse the first qudit with an ancilla of flux $b^{-1}$ and
trivial charge. The only state that can fuse into the vacuum with the ancilla
is $\ket{b}$, and this will happen with finite probability. Note that the
ancilla can never vanish into the vacuum with a state with charge because
there is no way of extending the basis to be invariant under the stabilizer 
group of the flux.

In the end, if the particles disappear into the vacuum, the ancilla is left
in the desired $X$ eigenstate. Otherwise, we repeat the procedure from the
beginning until eventually the state appears.

\subsection{Choosing a $d^{th}$ root of unity}

Before we continue building our gate-set, we have to address a problem
that appears for $d>2$, similar to the problem that occurred for
$d=2$ when proving that the gate-set is universal.

So far, we have defined everything in terms of $\omega$, a non-trivial $d^{th}$
root of unity. But there are $d-1$ of these, and there is a symmetry
which interchanges them. We will have to break this symmetry by using
an ancilla.

In particular, we need an ancilla that is an eigenstate of $X$ with 
eigenvalue not equal to $1$. We will then define this state to be
the $\ket{\tilde 1}$ state in the $X$ basis, i.e.,
\be
\ket{\tilde 1} = \frac{1}{\sqrt d} \sum_{i=0}^{d-1} \omega^{-i} \ket{i},
\ee
\noindent
which has eigenvalue $\omega$, thus fixing our root of unity. We then define
the other $X$ eigenstates by
\be
\ket{\tilde n} = \frac{1}{\sqrt d} \sum_{i=0}^{d-1} \omega^{-n i} \ket{i},
\ee
\noindent
and the operator $Z$ by 
$\ket{\tilde n} \rightarrow \ket{\widetilde{n-1}}$.

How do we produce the first $\ket{\tilde 1}$ in terms of which everything is
defined? We start with a $\ket{\tilde 0}$
(which is always well defined and which we know how to construct from 
the previous section), and we apply a controlled-$X^{-1}$ (which is a classical
function, and thus computable from the Toffoli) from this ancilla 
to a $\ket{0}$ ancilla, which produces the output
\be
\ket{\tilde 0}\otimes\ket{0} \longrightarrow 
\frac{1}{\sqrt{d}} \sum_i \ket{\tilde i} \otimes \ket{\tilde i}.
\ee

If we discard the second state, we will have a mixed state that is a 
combination of the different $X$ eigenstates. This is equivalent to being
handed an arbitrarily chosen $X$ eigenstate, which we will call 
$\ket{\tilde i}$.

We can obtain copies of this state by applying a controlled-$X^{-1}$
from a $\ket{\tilde 0}$ ancilla to this state, which applies the
transformation
\be
\ket{\tilde 0}\otimes\ket{\tilde i} \longrightarrow 
\ket{\tilde i}\otimes\ket{\tilde i}.
\ee
\noindent
We can thus build arbitrarily many copies of the state. We still have to
worry that this might be the $\ket{\tilde 0}$ state. However, below
in the section for measuring $X$, we will give a procedure to detect the 
$\ket{\tilde 0}$ which does not rely on having $\ket{\tilde 1}$ ancillas.
If we determine that $i=0$, we throw away all the copies and start over
(this will only happen with probability $1/d$). Otherwise,
we relabel our state as $\ket{\tilde 1}$, fixing a value for 
$\omega$.

Because we can copy the $\ket{\tilde 1}$ state, and below we will also
show how to measure it, we can build a reservoir of ancillas in this state,
which will be used for all future computations. We can even use copying,
comparing, and majority voting to error correct our reservoir, thus allowing
for computation even in the presence of noise.

\subsection{Measuring $X$}

The last gate needed for universality is the measurement of $X$. 
The basic idea is to fuse the pair of anyons that form the state
to be measured. The $\ket{\tilde 0}$ eigenstate will have some overlap with 
the vacuum, and will vanish with probability
$p=d/\lp|C(b)\rp|$, where $C(b)$ is the conjugacy class of $b$.

The other $X$ eigenstates have zero probability of vanishing because
$\ket{\tilde i} = \frac{1}{\sqrt{d}} \sum_i\omega^{-i} \ket{a^i b a^{-i}}$ 
is orthogonal to the vacuum for $i>0$.
To detect the state $\ket{\tilde i}$ we first apply a $Z^{i}$ and then use
the above fusion procedure. The $Z$ gate can be applied as a controlled-sum
with a $\ket{\tilde 1}$ target as discussed in Section~\ref{sec:gates}.

Of course, the above will require us to have many copies on which to measure,
which means we need to perform the transformation
\be
\sum_i C_i \ket{\tilde i} \longrightarrow \sum_i C_i 
\ket{\tilde i} \otimes  \ket{\tilde i} \otimes \ket{\tilde i} 
\otimes \cdots \otimes \ket{\tilde i},
\ee
\noindent
which is done using a controlled-$X^{-1}$ with a $\ket{\tilde 0}$ ancilla
as control and the state to be copied as target.

To perform the measurement non-destructively, we can fuse all but one of 
the copies of the state. Alternatively, using the $Z$ gate and 
$\ket{\tilde 0}$ ancillas, we can always produce the rest of the 
$\ket{\tilde i}$ states.
The rest of the logic is similar to the $Z$ measurement procedure. 

Having completed the construction of the universal gates, we have proven
that universal quantum computation can be performed with anyons from
simple and perfect finite groups. We now turn to the question of whether
these operations can be performed in a fault-tolerant fashion.

\section{\label{sec:leak}Leakage Correction}

In this section we will discuss both the motivation and the techniques needed
to implement error correction and fault tolerance in the software of an
anyonic computer. The main result will be the construction of a leakage 
correction circuit for anyons, which enables the use of the standard
techniques for handling errors.

\subsection{Motivation}

Any quantum system that uses non-locality to protect its data will
be susceptible to errors if a large number of its local components
are damaged simultaneously. The probability for failure is generally 
exponentially small in the size of the system, and is zero
in the theoretical limit of an infinite system. However, all physical systems
are finite. Furthermore, practical considerations may force a given setup
to have a size such that the error of probability is small but non-negligible.

In the case of anyons, errors can occur due to quantum tunneling, which is an
effect of the high-energy degrees of freedom that were frozen out to obtain
a two-dimensional discrete gauge theory. 
The probability of this type of error goes as
$e^{-m L}$, where $m$ is the mass of the lightest particle that can mediate a 
charge interaction, and $L$ is the separation between anyons. 

Finite temperature effects are another source of error. These effects involve 
the creation of charge pairs from the vacuum. Because these pairs have trivial
total charge, even if they braid with a computational anyon, the net
charges of the collective three particle excitation will still be correct.
However, if one of these particles separates from the group, or separately
braids with another anyon, then errors will be introduced. The density of the
thermal excitations goes as $e^{-\Delta / T}$, where $\Delta$ is the mass 
gap and $T$ is the temperature. 

A good anyonic quantum computer should therefore have $L>>m$ and $T<<\Delta$.
In some implementations, however, it may be more practical to accept a small 
error rate from the hardware, and then correct it using standard quantum error 
correction techniques. For such cases, we present below the necessary steps 
needed to implement software based error correction for anyons.

While any of the error correcting codes can be used, most techniques require 
embedding a code space inside a Hilbert space on which we can do universal 
quantum computation. However, in the case at hand, our computational states 
are embedded in a Hilbert space (the states with arbitrary flux and charge) 
in which we cannot perform universal quantum computation. Therefore, before
starting the recovery protocol, we must first deal with states that have 
leaked out of the computational subspace (the subspace in which we can perform 
universal computations).

\subsection{Implementation}

To deal with leakage errors we can construct a version of the 
swap-if-leaked gate described by Kempe \textit{et al}. \cite{Kempe:2001}.
The idea behind the gate is to implement a projective measurement
that can distinguish the computational subspace from its complement.
If a state if found to be in the computational subspace, it is left alone.
Otherwise, it is replaced with an arbitrary ancilla that is in the 
computational subspace. The ancilla will
still be an error, but one that is correctable by standard quantum
error correcting codes. In fact, the general methods of quantum error
correction and fault-tolerant computation can be applied to anyons
as long as we can reliably project leaked qudits into a state in the 
computational subspace.

We again focus on the case of simple and perfect groups, and defer the general
leakage correction protocol to the next section.
In the current formalism, the computational basis is the set of states of a 
pair of anyons with zero total magnetic charge, where each anyon
has zero electric charge and a magnetic flux of the form $a^i b a^{-i}$
or its inverse.

The first type of error that we will deal with, is when the total
magnetic flux of the pair is non-trivial. This is a particularly grievous 
error because, if we drag around a pair with a non-trivial net flux, we could
be introducing errors into all the other qudits. 
Furthermore, our assumption that we can perform the operation 
$h,g \rightarrow h,h g h^{-1}$ relied on the fact that the second
pair had zero net magnetic flux, so it is important that we detect and
fix this error first.

To detect a net flux, we take an ancilla $\ket{g}\otimes\ket{g^{-1}}$ and
encircle it by the qudit we are performing the leakage correction on.
The ancilla will get conjugated by the net flux of the qudit, and the 
qudit will get conjugated by the net flux of the ancilla which should be zero.
We then fuse the ancilla with a pair with opposite flux. If the net
flux of the qudit is in the stabilizer of $g$, the fusion will have vacuum
quantum numbers with a finite probability, whereas if the conjugation
changed the flux of the ancilla, there will always be a particle left behind.
If we repeat this many times with many different ancillas 
$\ket{g}\otimes\ket{g^{-1}}$, with good statistical confidence we will
be able to tell if the net flux of the qudit is in the stabilizer of $g$.
Because $G$ has no center, the intersection of all stabilizers is the
identity, and hence repeating the above with sufficiently many different 
elements $g$, we can detect a non-zero net magnetic charge.

If we detected a net flux, we replace the state with an ancilla in
the state $\ket{0}$. Of course, we must be very careful when moving the
damaged ancilla pair out of the region of qudits, so as not to damage
other states. That is, when moving
past other anyons, we always do so in the direction in which the
damaged pair gets conjugated and the good qudits are unaffected. 

In the case when the qudit passes the above test, then we have
projected into the zero net flux subspace, but otherwise left the state 
unchanged. The next step is to deal with electric charge.
Because it is very difficult to measure the electric charge of a single
anyon, we will start with a fresh ancilla $\ket{0}$, made from two
anyons neither of which have electric charge, and copy
the state over. Once again we will be using the incomplete swap circuit

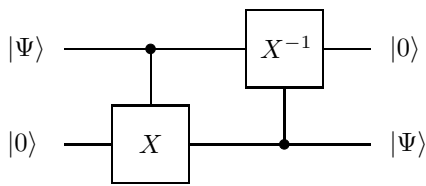
\begin{center}
\setlength{\unitlength}{1in}
\begin{picture}(3,1.1)(0,0)
{\put(.7,.8){\line( 1, 0){.95}}}
{\put(2.3,.8){\line( -1, 0){.25}}}
{\put(.7,.3){\line( 1, 0){.25}}}
{\put(2.3,.3){\line( -1, 0){.95}}}

{\put(1.15,.8){\line( 0,-1){.3}}}
{\thinlines \put(1.15,.8){\circle*{.05}}}
{\put(0.95,.1){\framebox(.4,.4){}}}

{\put(1.85,.3){\line( 0,1){.3}}}
{\thinlines \put(1.85,.3){\circle*{.05}}}
{\put(1.65,.6){\framebox(.4,.4){}}}

\put(1.15,.3){\makebox(0,0){$X$}}
\put(1.86,.81){\makebox(0,0){$X^{-1}$}}
\put(.4,.8){\makebox(0,0)[l]{$\ket{\Psi}$}}
\put(.4,.3){\makebox(0,0)[l]{$\ket{0}$}}
\put(2.4,.8){\makebox(0,0)[l]{$\ket{0}$}}
\put(2.4,.3){\makebox(0,0)[l]{$\ket{\Psi}$}}
\end{picture}
\end{center}

\noindent
when acting on the computational basis. Of course, the heart of a leakage 
detection algorithm is how to extend the operations outside of the 
computational subspace. The procedure cannot be described simply by 
a circuit, and therefore we will present a way of completing
the controlled-sum gate so that the above operation will always yield
a state that is in the computational subspace.

The following procedure is almost identical to the one used to produce
$\ket{\tilde 0}$ states. This is because $\ket{\tilde 0}$ states are obtained
by taking a vacuum state and projecting to the computational basis, which
is primarily leakage detection. The main difference is that when doing leakage
detection, we only get one chance of using the qudit (because of the no-cloning
theorem), but if the state leaked, it is acceptable to replace it by 
anything in the computational basis. The latter is clearly not acceptable 
when creating $\ket{\tilde 0}$ ancillas.

We will use the incomplete swap procedure for the second round of
leakage detection. Recall that by this point we have projected the qudit
into the zero net flux subspace. Take the qudit and a $\ket{0}$ ancilla,
and conjugate the ancilla by a function of the qudit's flux:
\be
f(a^i b a^{-i}) &=& a^i\nonumber\\*
f(\text{anything else}) &=& I.
\ee
\noindent
This is the same extension of a controlled sum that was used to produce
$\ket{\tilde 0}$ ancillas.

Afterward, we conjugate the original qudit by $f(g)^{-1}$, where $g$ is the 
flux of the ancilla. Note that because we know at this point that the
original qudit has net flux zero, the state of the ancilla will not exit
the computational basis during this operation
(though it might change within the computational basis
if the original state had non-zero electric charge). The result of the
past two controlled-sums is:
\be
&&\ket{\psi_\parallel}\otimes\ket{0} + \ket{\psi_\perp}\otimes\ket{0} 
\nonumber \\ &&\longrightarrow
\ket{0}\otimes\ket{\psi_\parallel} + 
\sum_{i=0}^{d-1} \ket{\psi_{\perp i}}\otimes\ket{i},
\ee
\noindent
where parallel and perpendicular refer to inside and outside the
computational basis, and none of the $\psi$ states are normalized.  
Finally, we replace the original pair with the ancilla pair and discard the 
original pair.

Clearly, the new state will be in the computational 
basis. Furthermore, if the original state was in the computational basis, 
then the new state will be equal to the old state, and unentangled
with the old anyons.

Having complemented our gate-set with a leakage correction scheme, we have 
proven not only that we can do universal quantum computation with anyons, 
but that these computations can be made fault-tolerant.

\section{\label{sec:hard}Universal Computation for Non-solvable Groups}

We will now generalize the results of the previous section to any non-solvable
group. Unfortunately, in our proofs for the simple perfect case,
we made extensive use of the fact that
we can compute any classical function simply by multiplying the inputs with
ancillas. This is no longer true, even if we restrict ourselves just to
perfect groups that are not simple. The quickest example is $A_5\times A_5$
which is perfect, but has two normal subgroups given by each of the $A_5$
factors.
Thus, if our two inputs are $1\times 1$ and $g\times 1$, there is no
expression made out of products in which the results differ in the second
factor.

The above example can easily be fixed by working within one $A_5$ subgroup.
In general, though, even this is not possible, as not all perfect groups have a
perfect and simple subgroup. However, the following theorem 
comes to the rescue:

\noindent
\textbf{Theorem:} If $G$ is a non-solvable finite group, then there exists a 
normal subgroup $P$ of $G$ and a subgroup $N$, normal in $P$, such that $P/N$
is perfect and simple.

\noindent
Once again we defer the proof to the appendix. 

What the theorem tells us
is that we want to work with cosets of $N$ in $P$. That is, we would like to 
replace our old flux eigenstates with states that are labeled by
elements in $P/N$ and invariant under $N$. A good guess would be
\be
\ket{x} = \frac{1}{\sqrt{\lp|N\rp|}} \sum_{n\in N} \ket{x' n},
\ee
\noindent
where $x$ is an element of $P/N$, and $x'$ is an element in the coset
that $x$ represents. More specifically, if $f:P\rightarrow P/N$ is the
canonical epimorphism that maps elements to cosets, then we require that 
$f(x')=x$. The particular choice of $x'$ has no effect on the above definition.

The above is a good guess but not quite right. A given coset may intersect
many different conjugacy classes of $G$, each of which lies in a different 
superselection sector. Thus, we are effectively working with mixed states. 

Remembering that we really want to keep our anyons in pairs of
zero net flux, the right choice for the new states is
\be
\rho_x = \frac{1}{\lp|N\rp|}\sum_{C\in {\mathcal C}(G)}\lp[
\lp( \sum_{x' \in (C \cap f^{-1}(x))} \ket{x'} \otimes \ket{x'^{-1}} \rp)\rp.
\ \ \ &&\\ \nonumber \otimes
\lp.\lp( \sum_{x' \in (C \cap f^{-1}(x))} \bra{x'} \otimes \bra{x'^{-1}} \rp)
\rp],&&
\ee
\noindent
where again $x$ is an element of $P/N$, and ${\mathcal C}(G)$ is the set
of conjugacy classes of $G$. 

These states have the nice property that when conjugated by any element
$h' \in P$ (or equivalently, when a flux $h'\in P$ is dragged around them), 
the effect 
only depends on the coset $f(h')$ of $h'$, and generates the 
transformation
\be
\rho_g \longrightarrow \rho_{f(h')\, g\, f(h')^{-1}}.
\ee 

Because of this, if we use the usual scheme of passing one pair of anyons
in between another, and they are both prepared in states of the above form,
the net effect is that the inner pair will get conjugated by the outer pair
as
\be
\rho_h\otimes \rho_g \longrightarrow \rho_h \otimes \rho_{h g h^{-1}},
\ee 
\noindent
keeping the pair unentangled.

\subsection{New requirements for the physical system}

While the operations of braiding, fusion, and vacuum pair creation
described in Section~\ref{sec:reqs}
all seem like reasonable requirements to demand from the physical
system, the requirement of flux ancillas is somewhat harder to justify.

In particular, take the case of a group that has a non-trivial center,
which can occur even if the group is perfect. Consider two fluxes $g$ and $c g$
that differ by multiplication of an element $c$ in the center. These two 
fluxes cannot be distinguished by conjugation, since 
$c g x (c g)^{-1}=g x g^{-1}$. Thus, it may be a difficult problem to distill
these flux eigenstates from the vacuum.

A more reasonable assumption is to require the existence of ancillas
only for the fluxes in the perfect subgroup. Another improvement might
be to assume that we only have ancillas in the mixed states $\rho_x$ defined
above, where $x\in P/N$. These states might be easier to produce because
they are obtained from the vacuum by first throwing away the anyons with 
flux not in $P$ or with non-trivial charge, and then projecting
to a definite coset of $N$ in $P$. Therefore, we will replace our old 
requirement for the existence of flux ancillas by:
\begin{enumerate}
\item[4'] We have ancillas in the state $\rho_x$ for any $x\in P/N$.
\end{enumerate}

It would be highly desirable to be able to prove that requirements 1 through
4 are sufficient to create the states in 4'. Unfortunately, it appears that 
requirements 1-3 combined with 4' may neither be a subset nor a superset
of requirements 1-4. Thus, in a sense, we are imposing a different set of 
requirements for this section. One ameliorating fact is that in
the case when $P$ is simple, the states $\rho_x$ are just flux 
eigenstates. We therefore could have used requirement 4' for all sections of 
this paper. We will not attempt to describe in the appendix a protocol by 
which these generalized ancillas can be created, however. 

\subsection{Universal computation}

As in Section~\ref{sec:basis}, we choose two elements $a,b\in P/N$ such that
$a^d b a^{-d}=b$ for some prime $d$, and $a^i b a^{-i} \neq b$ for $0<i<d$.
We then define our computational basis states as
\be
\rho_i = \rho_{a^i b a^{-i}},
\ee
\noindent
which we define as eigenstates of the $Z$ operator. The $X$ operator
is defined by the action $X(\rho_i) \equiv X \rho_i X^\dagger = \rho_{i+1}$
and its eigenstates can be obtained using the projection operators
$P_i = \sum_{j=0}^{d-1} \omega^{-i} X^i/d$ by
\be
\rho_{\tilde i} &=& d\times  P_i \rho_{0} P_i^\dagger \\\nonumber
&=& \frac{1}{d} \lp(\sum_{j=0}^{d-1} \omega^{-i} X^i\rp) \rho_0 
            \lp(\sum_{j=0}^{d-1} \omega^i X^{i\dagger}\rp).
\ee

At this point proving that universal quantum computation can be achieved
is fairly straightforward, and is almost identical to the discussion
in Section~\ref{sec:simp}. The major differences occur when we have to deal 
with states outside of the computational basis, that is, when creating 
$\rho_{\tilde 0}$ states and when dealing with leakage correction. Both
of these issues will be dealt with in the next subsection. As for the rest
of the operations, we will only give a very brief discussion:

Because the $\rho_x$ states have the same braiding properties as those of the
fluxes of a group $P/N$ (and in particular two $Z$ eigenstates remain
unentangled after braiding), the same method for producing a Toffoli 
applies to them.

Measuring $Z$ is easy because the $\rho_i$ states have support in orthogonal 
subspaces. The copy (using the Toffoli) and fuse with ancillas
procedure will work just as well as before.

For the interested reader, we will carry out below some of the calculations
needed to deal with $X$ eigenstates and prove universality. Most of the
results seem almost miraculous when expressed in the language of density
operators. However, the reader should bear in mind that we are only
using density operators to account for the different superselection sectors.
If we just fixed a superselection sector for each particle, we would be
dealing with pure states, and all of the proofs from the past section
would carry through.

We begin by studying the action of the controlled-$X^{-1}$ 
on $X$ eigenstates:
\be
&\rho_{\tilde m} \otimes \rho_{\tilde n}& =
\frac{1}{d} \sum_{i=0}^{d-1} \sum_{j=0}^{d-1} \omega^{-i m +j m} X^{i} \rho_0 
X^{j\dagger} \otimes \rho_{\tilde n}\nonumber\\
&\longrightarrow&
\frac{1}{d} \sum_{i=0}^{d-1} \sum_{j=0}^{d-1} \omega^{-i m+j m} X^{i} \rho_0 
X^{j \dagger} \otimes  X^{-i} \rho_{\tilde n} X^{-j \dagger}\nonumber\\
&=& \frac{1}{d} \sum_{i=0}^{d-1} \sum_{j=0}^{d-1} \omega^{-i m+j m}X^{i}\rho_0 
X^{j \dagger} \otimes  \omega^{-i n} \rho_{\tilde n} \omega^{j n}\nonumber\\
&=& \frac{1}{d} \sum_{i=0}^{d-1} \sum_{j=0}^{d-1} \omega^{-i+j (m+n)}
X^{i} \rho_0 X^{j \dagger} \otimes \rho_{\tilde n}\nonumber\\
&=& \rho_{\widetilde {m+n}} \otimes \rho_{\tilde n},
\ee
\noindent
which is equivalent to its action on pure states. 
Therefore, once we have $\rho_{\tilde 0}$ 
states, we can use the same trick as before to break the symmetry
and obtain a $d^{th}$ root of unity. That is, we do a 
controlled-$X^{-1}$ from the $\rho_{\tilde 0}$ with a $\rho_{0}$ target to 
create the state
\be
&\rho_{\tilde 0} \otimes \rho_{0}& \nonumber\\
&=&\lp(\frac{1}{d}\sum_{i=0}^{d-1}\sum_{j=0}^{d-1} 
X^{i}\rho_0 X^{j \dagger}\rp)
\otimes \lp(\sum_{n=0}^{d-1}\sum_{m=0}^{d-1} P_n \rho_0 P_m^\dagger\rp) 
\nonumber\\
&\longrightarrow&
\frac{1}{d}\sum_{i=0}^{d-1}\sum_{j=0}^{d-1} \sum_{n=0}^{d-1}\sum_{m=0}^{d-1}
X^{i}\rho_0 X^{j \dagger} \otimes \omega^{-i n+j m} P_n \rho_0 P_m^\dagger
\nonumber\\
&=& d \sum_{n=0}^{d-1}\sum_{m=0}^{d-1} \lp(P_n \rho_0 P_m^\dagger\rp)
\otimes \lp(P_n \rho_0 P_m^\dagger\rp)
\ee
\noindent
and then discard (trace out) the first state to get the state
$\rho = \sum_n P_n \rho_0 P_n = \sum_n \rho_{\tilde n} / d$, which gives us
an unknown eigenstate of $X$ as before. We then discard and repeat if we 
obtained the $\rho_{\tilde 0}$ state, and otherwise we relabel the state as
$\rho_{\tilde 1}$.

Once the $\rho_{\tilde 1}$ state is available, we can use a controlled-sum 
to produce the $Z$ gate, which will allow us to produce any $X$ ancilla
including more $\rho_{\tilde 1}$ states.

Finally, measuring $X$ works by fusing the pair of anyons, 
because the $\rho_{\tilde i}$ are orthogonal to the vacuum for $i>0$.
The full measurement proceeds as before by copying, permuting states
using the $Z$ gate, and then fusing.

\subsection{Leakage detection and $\rho_{\tilde 0}$ generation}

One final issue remains: How do we measure whether a state is in 
the computational subspace? Projecting onto the computational subspace
is useful because a $\rho_{\tilde 0}$ is just the projection of 
a vacuum state to the computational basis. Furthermore, this projection will
allow us to perform leakage correction.

One of the new issues that arises for general non-solvable groups is that
if we have a state in the computational basis, and we braid it with
an electric charge carrying a non-trivial representation of the subgroup
$N$, then the state will move outside the computational basis. The other
issue is that the conjugacy class of an element in $G$ might be larger
than the conjugacy class of the element in $P$, though given that
$P$ is normal, the first set will be entirely contained in $P$.

Let us begin by examining how the leakage correction algorithm must be 
changed. The first step
is to detect whether the net flux or charge of the pair of anyons
we are working on has a non-trivial effect on the states $\rho_x$.
The procedure is to braid the pair around the ancilla pair and then fuse
the ancilla with another ancilla in the state $\rho_{x^{-1}}$. If the
anyon pair has an effect on the ancilla states $\rho_x$, then the fusion
statistics will be altered, and this will be detectable after many repetitions.
If our state is found defective we discard it as usual, and replace it
by a state in the computational basis. Otherwise, we move on to the
next step. Note that if the anyon pair had a net flux in the subgroup
$N$, or in some element outside of $P$ that commutes with $P$, then
the state will still advance to the next round of error correction.
However, this anomalous flux or charge will not affect the usual braiding
properties.

The second round of error correction is a swap with an ancilla in the
$\rho_{0}$ state. Note that using our universal classical computation in 
$P$ we can guarantee that if the original state was in $P$, the ancilla
ends up in the computational basis. However, if the original anyons
are outside of $P$, we will get a state that is within $P$ (because $P$
is normal) but not necessarily in the computational subspace. The final 
step is to perform a swap with a second ancilla in the $\rho_{0}$
state, where now we know that the first ancilla had to be composed
of anyons with no charge, and fluxes only in $P$. This guarantees that the
final state of the second ancilla is in the computational basis, and equals 
the original state if it did not leak, completing the leakage correction
procedure.

To create $\rho_{\tilde 0}$ we also use a swap, this time between a pair
created from the vacuum and a $\rho_0$ ancilla. We then try to fuse the 
leftover vacuum state with a $\rho_{b^{-1}}$. If they fuse into the vacuum, 
then the ancilla is in a $\rho_0$ state. The logic is as follows: if the vacuum
pair had electric charge when created, then the swap will not change the
charge, and hence it cannot disappear into the vacuum. If the vacuum pair
has no electric charge but is outside of $P$, then the ancilla is still
guaranteed to be in $P$. Furthermore, when conjugating the vacuum state,
we will be conjugating by an element in $P$. The vacuum state will end
in a flux state outside of $P$, which is orthogonal to $\rho_{b^{-1}}$.
Finally, if the vacuum pair is a pair of fluxes in $P$, then it will be
of the form $\rho_{\tilde 0}$, possibly superposed with other states
$\rho_x$ outside the computational basis. But the generalized swap
can guarantee that a state in $P$ outside of the computational
basis, will remain outside of the computational basis (just like in the
simple perfect case). Only when the ancilla is in the
state $\rho_{\tilde 0}$ can the fusion into the vacuum occur.

The above procedure for producing $\rho_{\tilde 0}$ ancillas completes
the gate-set for non-solvable groups, and proves the main result of this
paper: that anyons with fluxes
in a non-solvable group can perform universal quantum computation.

\section{Conclusions and Outlook}

While we have shown that universal quantum computation is theoretically 
feasible for any non-solvable group, it is still not yet clear whether we
will ever be able to build an anyon based computer. First of all there
is the fact the smallest non-solvable group is $A_5$ which has 60 elements.
Obtaining such a group from symmetry breaking seems problematic.

One may wonder whether we can do computation with solvable groups.
For abelian groups, each superselection sector consists of just one state,
so it is not possible to encode quantum data in a topologically invariant 
fashion. 
Attempts such as Ref.~\cite{Lloyd} encode the quantum data in a superposition
of position eigenstates, but this has no more robustness than using
superpositions of positions of any other neutral particle of
the same mass. 

Hope still remains for solvable non-abelian groups. While producing
Toffoli gates using conjugation as in this paper will most likely no longer 
be feasible, Toffoli gates might still be performed by employing magic states.
In fact, Kitaev has such a procedure for the group $S_3$ \cite{Kitaev:S3}.
The full set of groups which can perform universal quantum computation 
remains unknown, but we believe it does not include every non-abelian 
group. 

Furthermore, there are anyons that are not based on the electric and magnetic 
charge model (quantum double of a group) presented here. 
Some of the more exotic
anyons are likely to be good quantum computers, but in general, their
computational power remains unknown.

We have also neglected to present in this paper an account of the resources
used to perform computations. While it should be clear that computations
can be done with at worst a polynomial overhead in the size of the input,
some gates (in particular those that require calculations of arbitrary 
functions over the group) may require resources that are exponential in the
size of the group. A lot of the wasted resources may come from the
description in terms of general groups, though. For a fixed group, 
the resources can probably be significantly reduced.

Finally, there remains the question of physical systems which contain anyons.
Because of the requirement of two dimensions, we must look for quasi-particles
in some two dimensional medium. There are some indications that non-abelian
anyons may arise in the fractional quantum Hall effect (see 
Refs.~\cite{Ogburn:1998,Preskill:1997uk} and references therein). 
However, at the moment,
there are no physical systems out of which the anyonic computer may be built.
Even if no physical implementations are ever found, though, this subject
will hopefully still be interesting because of its beautiful mix of
computation, particle physics, and group theory.

\begin{acknowledgments}

The idea for universal classical computation with simple and perfect groups
was initially suggested by Alexei Kitaev, to whom I am highly grateful. 
The author would also like to thank John Preskill, Jim Harrington, and 
James Chakan for proofreading this paper, and Meg Wessling both
for translating this paper into proper English and for many 
useless but entertaining conversations.

This work was supported in part by
the National Science Foundation under grant number EIA-0086038 and
by the Department of Energy under grant number DE-FG03-92-ER40701.

\end{acknowledgments}

\appendix
\section{\label{sec:ancillas}Creating the ancillas}

As discussed in the main text, the requirement of a supply of calibrated flux 
ancillas needs further justification. In this section we will
show that for a perfect and simple group, the requirements of braiding,
fusion, and vacuum pair creation can be supplemented by one extra measurement
to allow the distillation of flux ancillas. We will not cover the general
non-solvable case, though.

The new measurement involves determining whether a single anyon has
trivial flux or not. Indeed, this measurement may even be done destructively.
The plausibility of this measurement relies on the fact that non-zero
flux charges are topologically non-trivial configurations that often
have much higher masses than their electric charge counterparts.
Naturally, dyons also have large masses and will be detected as having
non-trivial flux.

\subsection*{Step 1: Creating electric ancilla pairs}

The procedure for creating flux ancillas begins by creating single anyons 
with zero flux. 
These are obtained by creating a vacuum pair, measuring the
flux of the first particle of the pair, and discarding the second one if 
the first one had non-trivial flux.

The next step is to create pairs of anyons, where each anyon
has zero flux and unknown charge, but the pair has vacuum quantum numbers.
Of course, if we could non-destructively distinguish trivial from nontrivial 
flux, we could skip this step, as the vacuum pairs always have vacuum quantum 
numbers.

Take two of the single electric charges we have produced. We are going to
try to project this state onto the desired state with vacuum quantum numbers. 
Consider the process of creating a pair of anyons from the vacuum, braiding
one of them around the pair of charges, and then fusing the vacuum pair.
If the pair of charges had vacuum quantum numbers, then the vacuum pair
will remain in the vacuum state throughout this process, and fuse into the
vacuum at the end with unit probability. On the other hand, if the pair of 
charges does not have vacuum quantum numbers, then there will be a finite 
probability that the pair created from the vacuum will leave a particle behind
after fusion (since the vacuum is the only state that is left invariant by
the action of every flux).

Repeated application of this process will be a projective measurement 
which determines
whether the pair of charges has vacuum quantum numbers. If we project onto a
vacuum pair, then we have found a good charge ancilla pair. If the pair
does not project onto the vacuum state (because the two anyons 
do not transform 
in conjugate representations, or because we projected to a state orthogonal
to the vacuum), then we pair them up with other charges and repeat the process.
While slow, this process will eventually yield as many electric charge
pairs with vacuum quantum numbers as needed.

\subsection*{Step 2: Identifying the magnetic charges}

The electric ancilla pairs are useful because they can perform
a non-destructive measurement of magnetic flux. The procedure is
to take a member of the electric charge pair, drag it around the 
anyons or group of anyons whose total flux we want to measure,
and then fuse it with its pair. 

To describe the effect of the fluxes, let $R(g)$ be the representation of the
first electric charge of the pair. Let $\ket{n}$ be an orthonormal basis
for the space on which $R$ acts, and let $\ket{n^{*}}$ be the dual basis
for the conjugate representation $R^*$ under which the second charge
transforms. The effect of a flux $g$ is then
\be
\sum_n \ket{n} \otimes \ket{n^{*}} \longrightarrow 
\sum_n \lp( R(g) \ket{n}\rp) \otimes \ket{n^{*}}.
\ee

Just as before, if the total flux is non-trivial, there will be a good 
chance that the fusion of the electric charges will leave
a particle behind. On the other hand, if the total flux is trivial,
even if the total charge is not, the pair of electric charges will always fuse
into the vacuum.

Repeated application of this procedure will determine
whether the total flux is trivial or not. Furthermore, this procedure 
will at worst introduce decoherence in the flux basis, but will leave all
flux eigenstates unchanged. 

\begin{figure}
\setlength{\unitlength}{0.00083333in}
{\renewcommand{\dashlinestretch}{30}
\begin{picture}(3042,775)(0,-10)
\put(2933,305){\blacken\ellipse{150}{150}}
\put(2933,305){\ellipse{150}{150}}
\put(2483,305){\blacken\ellipse{150}{150}}
\put(2483,305){\ellipse{150}{150}}
\put(1733,305){\blacken\ellipse{150}{150}}
\put(1733,305){\ellipse{150}{150}}
\put(1283,305){\blacken\ellipse{150}{150}}
\put(1283,305){\ellipse{150}{150}}
\put(83,305){\ellipse{150}{150}}
\put(383,305){\ellipse{150}{150}}
\path(83,155)(84,154)(87,152)
	(92,149)(101,144)(112,138)
	(126,130)(144,122)(163,113)
	(185,103)(209,94)(234,85)
	(260,77)(287,69)(316,62)
	(346,56)(378,50)(413,45)
	(450,41)(491,37)(535,33)
	(583,30)(613,28)(644,27)
	(676,25)(710,23)(745,22)
	(781,21)(819,19)(859,18)
	(899,17)(941,16)(984,15)
	(1029,15)(1074,14)(1120,13)
	(1167,13)(1215,12)(1263,12)
	(1311,12)(1360,12)(1408,12)
	(1457,12)(1505,12)(1553,13)
	(1600,13)(1647,14)(1692,15)
	(1737,15)(1780,16)(1823,17)
	(1864,18)(1904,19)(1943,21)
	(1980,22)(2016,23)(2050,25)
	(2083,27)(2115,28)(2146,30)
	(2191,33)(2234,36)(2275,40)
	(2313,44)(2349,48)(2383,53)
	(2415,58)(2445,64)(2473,70)
	(2498,76)(2522,83)(2544,90)
	(2563,98)(2580,106)(2596,115)
	(2610,124)(2622,133)(2632,142)
	(2641,152)(2649,162)(2655,172)
	(2661,183)(2666,194)(2671,205)
	(2676,221)(2681,237)(2685,255)
	(2689,274)(2692,294)(2694,314)
	(2696,336)(2697,358)(2698,380)
	(2697,402)(2696,424)(2694,446)
	(2692,466)(2689,486)(2685,505)
	(2681,523)(2676,539)(2671,555)
	(2665,568)(2659,581)(2652,594)
	(2645,606)(2636,618)(2627,629)
	(2616,640)(2605,650)(2592,660)
	(2578,670)(2564,678)(2548,687)
	(2531,694)(2514,701)(2496,707)
	(2478,713)(2458,718)(2438,722)
	(2417,726)(2396,730)(2375,733)
	(2353,736)(2330,738)(2305,740)
	(2280,742)(2253,744)(2226,745)
	(2197,747)(2168,747)(2138,748)
	(2108,748)(2078,748)(2048,747)
	(2019,747)(1990,745)(1963,744)
	(1936,742)(1911,740)(1886,738)
	(1863,736)(1841,733)(1821,730)
	(1799,726)(1778,722)(1758,718)
	(1739,713)(1720,707)(1702,701)
	(1685,694)(1669,686)(1654,677)
	(1640,668)(1626,658)(1614,647)
	(1604,636)(1594,624)(1586,612)
	(1578,599)(1572,586)(1566,572)
	(1562,558)(1558,542)(1555,525)
	(1552,506)(1550,486)(1548,465)
	(1546,442)(1545,419)(1543,395)
	(1542,370)(1540,346)(1538,322)
	(1536,298)(1533,275)(1529,254)
	(1524,233)(1518,214)(1512,197)
	(1504,182)(1496,167)(1486,156)
	(1476,146)(1464,136)(1451,128)
	(1437,120)(1422,114)(1405,108)
	(1388,103)(1370,100)(1351,97)
	(1332,96)(1313,95)(1293,95)
	(1274,97)(1255,98)(1237,101)
	(1220,104)(1203,108)(1186,113)
	(1171,117)(1154,124)(1137,130)
	(1121,138)(1104,146)(1088,154)
	(1072,163)(1055,173)(1039,183)
	(1022,193)(1006,202)(990,212)
	(974,222)(958,231)(943,239)
	(928,247)(913,255)(898,261)
	(883,267)(867,273)(851,278)
	(834,282)(815,286)(795,289)
	(773,292)(748,294)(721,297)
	(692,299)(662,300)(632,302)
	(604,303)(579,304)(559,304)(533,305)
\blacken\path(654.064,330.366)(533.000,305.000)(651.758,270.410)(654.064,330.366)
\put(83,455){\makebox(0,0)[b]{$e$}}
\put(383,455){\makebox(0,0)[b]{$\bar e$}}
\put(1208,455){\makebox(0,0)[lb]{$g_2$}}
\put(1658,455){\makebox(0,0)[lb]{$g_2^{-1}$}}
\put(2408,455){\makebox(0,0)[lb]{$g_1$}}
\put(2858,455){\makebox(0,0)[lb]{$g_1^{-1}$}}
\end{picture}
}
\caption{\label{measure}Using electric charges to check if $g_1=g_2$.}
\end{figure}
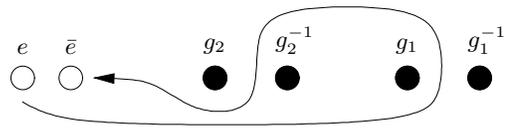

We can use this procedure to compare the fluxes of two anyons. In particular,
consider two pairs created from the vacuum. Measure the total flux
of the first anyon of the first pair combined with the second anyon of the
second pair. If the combined flux is trivial, the first anyon of each pair
has the same flux; otherwise the flux is different. The procedure
is depicted in Figure~\ref{measure}.

The above procedure allows us to sort the flux pairs into ``bins'' that
depend on the total flux of the first anyon of the pair. We will get as
many bins as elements of $G$, each containing an unlimited supply
of vacuum pairs which carry the same flux in the first
anyon of the pair. At this point, if the fluxes have not decohered in the
flux basis, then we must have an entangled state involving all anyons in
a given bin. Throwing away a single flux from each bin will produce
the desired decoherence, just as it did when breaking the
various symmetries in the main part of this paper.

All that remains is to identify each bin with an element of $G$. 
Assume that we were given an assignment of an element of $G$ to each bin.
The assignment could be checked by using the following procedure.
First, we note that any finite group $G$ may be described by a set of
elements $\lp\{g_i\rp\}$ and a set of relations of the form
$g_{i_1}\cdots g_{i_n}=1$ which they obey. To check that the assignment
is correct, we just need to check all the relations (supplemented by the
trivial one element relations $g_i^n g_i^m = g_i^{n m}$). These can
be checked again with the electric charge ancillas, using a loop that 
circles each of the fluxes in the relation in the correct sequence.

To generate guesses, we could just randomly assign to each bin an element
of $g$, which gives us a probability of success of at least $1/(|G|)!$.
Of course, we can be a lot smarter, as the above procedure can
help us figure out the powers of a given element (including its inverse)
and even the elements in its conjugacy class. Thus the need for guesswork is
minimal, and some of the choices correspond to different valid assignments
(i.e., automorphisms) of the group.

\subsection*{Analysis of the produced ancillas}

At this point we have almost produced the desired ancillas, with one caveat:
the individual anyons do not have trivial charge (i.e., they may be dyons).
However, all we have done to the pairs, after creating them from the vacuum,
is to measure the flux of one of the anyons. That means that the electric
charge portion of the state is still in the vacuum state. More technically,
if the ancilla pair circles a flux that commutes with the flux of one
anyon of the ancilla, then the state remains unaltered. This is the same
behavior that the pure magnetic charges would have.

Some careful thought at this point shows that these states are good enough
for the quantum computation procedure presented in the bulk of the paper.
Indeed, going back and repeating all the steps with these generalized ancillas
would require very few modifications. The fusion to measure in the $Z$ basis 
would now have a lower success probability, which is compensated by a higher 
rate of producing acceptable $\ket{\tilde 0}$ states, but otherwise most 
gates remain unaltered. 
We have therefore succeeded in constructing an ancilla reservoir, 
which, while slightly different then the one initially 
desired, is useful for universal quantum computation.

\section{\label{sec:proofs}Mathematicalia}

This appendix proves the major mathematical theorems needed in the bulk
of the paper. We begin by stating the definitions of some of the 
mathematical terms used:

\noindent
\textit{Perfect group}:
A non-trivial group G such that $[G,G]=G$. 
Note that $[G,G]$ is not the set of elements
of the form $[g_1,g_2]\equiv g_1 g_2 g_1^{-1} g_2^{-1}$ 
but rather the group generated by these elements.
Even if $G$ is perfect, there may not be a commutator expression for every
element.

\noindent
\textit{Non-solvable group}:
A group that has a perfect subgroup.

\noindent
\textit{Normal subgroup}: 
A subgroup $H$ of a group $G$ such that $g h g^{-1}\in H$
for every $h\in H$ and $g\in G$.

\noindent
\textit{Simple group}: 
A group with no normal subgroups other than the whole group
and the trivial group.

Before we get to our main theorem, we will prove a theorem that will allow 
us to deal with general non-solvable groups. We intend to show that we can 
extract from non-solvable groups a simple and perfect group.
The simple perfect groups (which can also be described as the simple
non-abelian groups) are the ones on which we can perform universal classical
computation and are therefore important for this paper.

We begin by defining the $n^{th}$ derived subgroups by the
relations $G^{(n)}=[G^{(n-1)},G^{(n-1)}]$ and $G^{(1)}=[G,G]$.
A solvable group is one for which $G^{(i)}=\{1\}$ for some $i$.
A non-solvable group must have an $i$ such that for every $j>i$, 
$G^{(i)}=G^{(j)}$ and $G^{(i)}$ is non-trivial. The group $G^{(i)}$ is 
perfect, thus the definition for solvable groups is
consistent with the definition for non-solvable groups given above.

Furthermore, all the groups $G^{(n)}$ are normal subgroups of $G$. 
This can be proven by recalling the property 
$g[g_1,g_2]g^{-1}= [g g_1 g^{-1},g g_2 g^{-1}]$. The rest follows by
induction because $G^{(1)}$ is normal in $G$, and $G^{(i)}$ is normal in $G$ 
if $G^{(i-1)}$ is. We have therefore shown that every non-solvable group
$G$ has a perfect normal subgroup $P$.

Sadly, this subgroup is not necessarily simple. However, we can prove
that every perfect group $P$ has a normal subgroup $N$ such that 
$P/N$ is perfect and simple. We choose $N$ to be a normal proper subgroup of 
$P$ such that no other normal proper subgroup of $P$ has more elements, which 
is well defined because $P$ is finite. Let $f$ be the canonical epimorphism 
$P \rightarrow P/N$ which maps elements into cosets. Because $f$ is
surjective we have $[P/N,P/N] = [f(P),f(P)] = f([P,P]) = f(P) = P/N$,
which, combined with the fact that $P/N$ is non-trivial, shows that
$P/N$ is perfect.

Finally, assume that $P/N$ has a normal, non-trivial proper subgroup
$A$. Then $B=f^{-1}(A)$ is a normal subgroup of $P$, because for 
any elements $b_1, b_2 \in B$ and $p \in P$, we have 
$f(b_1 b_2) = f(b_1) f(b_2) \in A$ and 
$f(p b_1 p^{-1}) = f(p) f(b_1) f(p)^{-1} \in A$.
Furthermore, $B$ is a proper subgroup of $P$, and 
$N = f^{-1}(1)$ is smaller than  $B = f^{-1}(A)$, leading to a contradiction.
Therefore, $P/N$ is simple, and we have finished proving the following
theorem:

\noindent
\textbf{Theorem:} If $G$ is a non-solvable finite group, then there exists a 
normal subgroup $P$ of $G$ and a subgroup $N$, normal in $P$, such that $P/N$
is perfect and simple.

\noindent
\textbf{Proof:} Shown by the above text.

We now turn our attention to using our groups to compute classical functions.
We shall prove that the set of functions that can be written in product form 
is complete, in the sense that it includes every function from 
$G^n\rightarrow G$,
if $G$ is simple and perfect (or equivalently simple and non-abelian).
This was first proven in the mathematical literature by Maurer in 1965
\cite{Maurer:1965}. In the computer science literature, a related result
was proven by Barrington \cite{Barrington:1989}.
In this paper, we will provide our own constructive 
proof for the following theorem:

\noindent
\textbf{Theorem:} If $G$ is a simple and perfect finite group, 
then any function
$f(g_1,\dots,g_n):G^n\rightarrow G$ can be expressed as a product of the
inputs $\{g_i\}$, their inverses $\{g_i^{-1}\}$ and fixed elements of $G$, 
any of which may appear multiple times in the product.

\noindent
\textbf{Proof:} 
Throughout this proof we will refer to the set of functions 
that can be expressed in the above form as ``computable.'' Proving the
above statement is equivalent to showing that all functions are computable.
The proof consists of building a series of computable delta functions that
map most elements to the identity, and then expressing arbitrary functions
as a product of these delta functions.

\noindent
\textbf{Step 1:} Given a group element $a$ not equal to the identity, 
let $C(a)$ denote its conjugacy class. Then the subgroup generated 
by the elements of $C(a)$ is equal to $G$. This is because the
subgroup is a nontrivial, normal subgroup of $G$ and $G$ is simple.

\noindent
\textbf{Step 2:} Fix two disjoint subsets $A$ and $B$ of $G$. Define a family 
of functions $\{\delta_c^{A,B}(g):A \cup B\rightarrow G\}$ with elements 
labeled by $c\in G$:
\be
\delta_c^{A,B}(a)=1& \ \ \ &\forall \ a\in A\nonumber\\
\delta_c^{A,B}(b)=c& \ \ \ &\forall \ b\in B.
\ee
\noindent
If the function $\delta_c^{A,B}$ is computable for some $c\neq 1$, then every 
function in the family is computable. To prove this choose any $d\in G$. 
By Step 1 there is an expression for $d$ as a product of elements in
the conjugacy class of $c$ (for instance, 
$d = g_1 c g_1^{-1} g_2 c g_2^{-1} c$).
Then $\delta_d^{A,B}$ is obtained by substituting $\delta_c^{A,B}$ for $c$ 
in the expression.

\noindent
\textbf{Step 3:} Fix a set $A$, an element $b$ not in $A$, and 
an element $x\neq b$. If a function $\delta_c^{A,B}$ is computable for some 
$B$ such that $b\in B$, then there exists a computable function 
$\delta_c^{A',B'}$ with two new sets such that
$A \cup \{x\} \subset A'$ and $b\in B'$. The function can be obtained from
\be 
\delta_e^{A',B'}(g) = [\delta_d^{A,B}(g),g x^{-1}]
\ee
\noindent
using Step 2. The above equation assumes that we have extended the domain 
of $\delta_d^{A,B}$ to $G$, which can be done in a natural way once we have
fixed a product representation for $\delta_d^{A,B}$.
The element $d$ was chosen to not commute with $b x^{-1}$. Such an element
must exist because $G$ is simple and non-abelian, and hence has no center. 
The element $e$ is just $e=[d,b x^{-1}]$.

\noindent
\textbf{Step 4:} The functions defined by 
$\delta_c^b(g)\equiv \delta_c^{A,B}$, with $A = G-\{b\}$
and $B = \{b\}$, are computable. To prove this start with $A_1=\{1\}$ and 
$B_1=\{b\}$. The function $\delta_c^{A_1,B_1}$ is computable because it is 
in the same family as $f(g)=g=\delta_g^{A_1,B_1}$. Then proceed by 
induction, using Step 3, on the elements in $G-\{b\}$ that are not 
included in $A_i$.  

\noindent
\textbf{Step 5:} For a fixed set of ordered elements $b_1,\dots,b_i$ 
define a family of functions labeled by $c$:
\be
\delta_c^{b_1\dots b_i}\lp(g_1,\dots,g_i\rp) = c&\ \ \ \ &
g_1=b_1,\dots, \text{and}\  g_i=b_i \nonumber\\
\delta_c^{b_1\dots b_i}\lp(g_1,\dots,g_i\rp) = 1&\ \ \ \ &\text{otherwise}.
\ee
\noindent
The same proof in Step 2 shows that if any function of the family with 
$c\neq 1$ is computable, then the entire family is computable.

\noindent
\textbf{Step 6:} Fix $i\in Z^+$ and elements $b_1,\dots,b_{i+1}\in G$.
If the function $\delta_c^{b_1\dots b_i}(g_1,\dots,g_i)$ is computable, 
then so is the function $\delta_c^{b_1\dots b_{i+1}}(g_1,\dots,g_i,g_{i+1})$. 
By Step 5 it is sufficient to be able to compute
\beq
\delta_e^{b_1\dots b_{i+1}}(g_1,\dots,g_{i+1}) = 
\lp[\delta_c^{b_1\dots b_i}(g_1,\dots,g_i),\delta_d^{b_{i+1}}(g_{i+1})\rp],
\eeq
\noindent 
where the function $\delta_d^{b_{i+1}}(g_{i+1})$ is computable by Step 4, 
and $d$ is chosen so that $e=[c,d]\neq 1$.

\noindent
\textbf{Step 7:} Using induction on the number of inputs of the function, and
starting from the base case $\delta_c^{b_1}(g_1)$, it is clear that all the 
functions defined in Step 5 are computable.

\noindent
\textbf{Step 8:} Every function is computable because
\be
f\lp(g_1,\dots,g_i\rp) = \prod_{b_1\in G}\cdots\prod_{b_i\in G}
\delta_{f\lp(b_1,\dots,b_i\rp)}^{b_1\dots b_i}\lp(g_1,\dots,g_i\rp). 
\nonumber\\
\mbox{\textbf{\ Q.E.D.}} \nonumber
\ee


\end{document}